\documentclass[useAMS,usenatbib]{mnras}

\usepackage[english]{babel}
\usepackage{amsmath,amssymb}
\usepackage{graphicx}
\usepackage{subfig}
\usepackage[normalem]{ulem}

\usepackage{verbatim}

\usepackage{color}
\usepackage{multirow}
\usepackage{mathtools}
\usepackage{epstopdf}

\usepackage{url}
\usepackage{color}

\def\mnras{MNRAS}
\def\apj{ApJ}
\def\apjl{ApJL}
\def\apjs{ApJS}
\def\aap{A\&A}
\def\aaps{A\&A Supp.}

\def\araa{ARA\&A}

\def\apss{Ap\&SS}               

\def\pasp{Publ. Astr. Soc. Pac.}

\def\be{\begin{equation}} 
\def\ee{\end{equation}} 
\def\ba{\begin{eqnarray}} 
\def\ea{\end{eqnarray}}

\def\cc{\,{\rm {cm^{-3}}}} 
\def\msun{{\Msun}}

\def\gsim{\lower.5ex\hbox{\gtsima}} 
\def\lsim{\lower.5ex\hbox{\ltsima}} \def\gtsima{$\; \buildrel > \over 
\sim \;$} \def\ltsima{$\; \buildrel < \over \sim \;$} \def\prosima{$\; 
\buildrel \propto \over \sim \;$} \def\gsim{\lower.5ex\hbox{\gtsima}} 
\def\lsim{\lower.5ex\hbox{\ltsima}} 
\def\simgt{\lower.5ex\hbox{\gtsima}} 
\def\simlt{\lower.5ex\hbox{\ltsima}} 
\def\simpr{\lower.5ex\hbox{\prosima}} \def\la{\lsim} \def\ga{\gsim} 
  
 \def\gtsima{$\; \buildrel > \over \sim \;$} 
\def\ltsima{$\; \buildrel < \over \sim \;$} 
\def\gsim{\lower.5ex\hbox{\gtsima}} 
\def\lsim{\lower.5ex\hbox{\ltsima}} 
\def\simgt{\lower.5ex\hbox{\gtsima}} 
\def\simlt{\lower.5ex\hbox{\ltsima}} 
\def\simpr{\lower.5ex\hbox{\prosima}} 
\def\la{\lsim} 
\def\ga{\gsim}

\def\msun{\,{\rm \Msun}}

\def\E3{{\cal E}_{\rm g}^{III}}

\def\r12{r_{1/2}} 
\def\x12{x_{1/2}} 
\def\v12{v_{1/2}}

\def\msun{{\rm M}_{\odot}}
\def\zsun{{\rm Z}_{\odot}}

\newcommand\textlcsc[1]{\textsc{\MakeLowercase{#1}}}
\newcommand{\quotes}[1]{``#1''}
\newcommand{\quotesing}[1]{`#1'}

\def\angstrom{\textrm{A\kern -1.3ex\raisebox{0.6ex}{$^\circ$}}}

\usepackage[all]{hypcap} 

\def\ccol{{\rm cm}^{-2}}

\newcommand\altaffilmark[1]{$^{#1}$}

\title[Physical properties of galaxies from their spectra]{Inferring physical properties of galaxies from their emission line spectra}

\author[Ucci G. et al.]{
	Ucci G.\altaffilmark{1}\thanks{E-mail: graziano.ucci@sns.it}, Ferrara A.\altaffilmark{1,2}, Gallerani S.\altaffilmark{1}, Pallottini A.\altaffilmark{1,3,4}
	\\
	$^1$Scuola Normale Superiore, Piazza dei Cavalieri 7, 56126, Pisa, Italy\\
	$^2$Kavli IPMU, The University of Tokyo, 5-1-5 Kashiwanoha, Kashiwa 277-8583, Japan\\
	$^3$Cavendish Laboratory, University of Cambridge, 19 J. J. Thomson Ave., Cambridge CB3 0HE, United Kingdom\\
	$^4$Kavli Institute for Cosmology, University of Cambridge, Madingley Road, Cambridge CB3 0HA, UK}

\date{Accepted XXX. Received YYY; in original form ZZZ}

\pubyear{2016}

\begin{document}
\label{firstpage}
\pagerange{\pageref{firstpage}--\pageref{lastpage}}
\maketitle

\begin{abstract}
We present a new approach based on Supervised Machine Learning (SML) algorithms to infer key physical properties of galaxies (density, metallicity, column density and ionization parameter) from their emission line spectra. We introduce a numerical code (called \textlcsc{GAME}, GAlaxy Machine learning for Emission lines) implementing this method and test it extensively. \textlcsc{GAME} delivers excellent predictive performances, especially for estimates of metallicity and column densities. We compare \textlcsc{GAME} with the most widely used diagnostics (e.g. R$_{23}$, [NII]$\lambda$6584 / H$\alpha$ indicators) showing that it provides much better accuracy and wider applicability range. \textlcsc{GAME} is particularly suitable for use in combination with Integral Field Unit (IFU) spectroscopy, both for rest-frame optical/UV nebular lines and far-infrared/sub-mm lines arising from Photo-Dissociation Regions. Finally, \textlcsc{GAME} can also be applied to the analysis of synthetic galaxy maps built from numerical simulations.
\end{abstract}


\section{Introduction}
Most of the information on the physical properties of galaxies is encoded in their spectra. These are characterized by a large number of emission lines from which their internal structure can be inferred. Several attempts have been made to recover such physical properties by mean of diagnostics based on small, selected subsets of emission line ratios. Most of these previous works focused on ionized nebulae and have obtained diagnostics for the physical properties of galaxies based only on the strongest nebular emission lines coming from extra-galactic HII regions and star-forming galaxies.
	
In principle once calibrated, a diagnostic should be a univocal function of a given physical parameter. For example, each value of the R$_{23}$ diagnostic (see Sec. \ref{sec:overview}) should trace a particular value of the metallicity of the gas. However, one of the most limiting aspects of this approach is that it suffers from many systematic errors which may plague the calibration, i.e. non-monotonic behavior, sizable scatter in the calibration and also that most diagnostics are only limited to Strong Emission Lines \citep{Lòpez2012}. Finally, in galaxies the Interstellar Medium (ISM) is characterized not only by nebular emission lines typical of HII regions tracing the metallicity, the ionization parameter or the temperature of the gas \citep{Kewley2001,Kewley2002,Levesque2010} but also by lines arising from low-ionization species (e.g. [CI], [OI], [NI] and [SII]) typical of neutral and denser regions.
	
A relatively new interesting field of research in astrophysics is represented by Machine Learning (ML) methods. These have already provided, for example, excellent predictive accuracy in the determination of photometric redshifts or in the star-galaxy classification task \citep{Ball2006,Kim2015}. A number of numerical techniques have been applied such as Artificial Neural Networks \citep{Collister2004} or Decision Trees \citep{Carliles2010,Carrasco2013,Cavuoti2015}. For a review of the applications of ML to astrophysical problems we refer the reader to \citet{Ball2010} and \citet{Ivezic2014}. For other practical applications of these methods see also \citet{Ball2008,Hoyle2015a,Zitlau2016,Hoyle2015b,Bellinger2016} and \citet{Kamdar2016} for a ML framework applied to cosmological simulations and \citet{Jensen2016} for a ML approach to measure the escape fraction from galaxies in the EoR (Epoch of Reionization).
	
The main purpose of this work is to reconstruct key physical parameters of distant galaxies (metallicity, column density, ionization parameter, density) once their spatially resolved spectra have been acquired. Our aim is to maximize the information that can be extracted from such data by using not only few specific and pre-selected emission lines, but the full information encoded in the spectra. This is now possible thanks to the combination of powerful Supervised Machine Learning algorithms and large synthetic spectra libraries. As we will see in the following, we covered a very large range of plausible physical properties of ISM clouds to accurately describe the physics beyond not only the ionized regions but also of other phases (i.e neutral, molecular) of the ISM in galaxies.

\begin{figure*}
	\centering
	\includegraphics[width=0.9\linewidth]{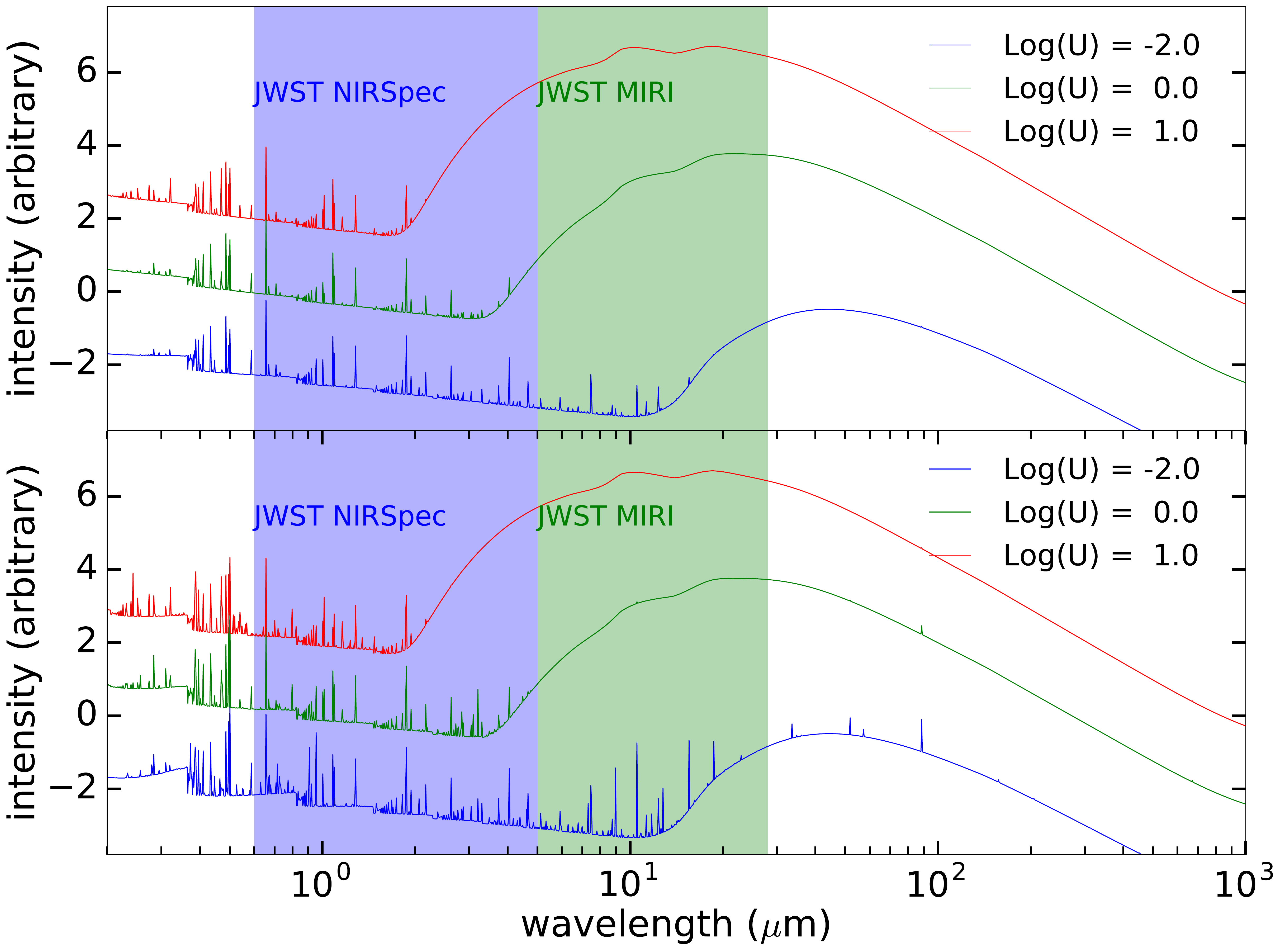}
	\caption{Rest-frame spectra obtained from our grid of models with metallicity $Z = 0.005 \zsun$ (\textit{upper panel}) and $Z = 0.5 \zsun$ (\textit{lower panel}). In both panels the density and column density are: $\log(n/\cc) = 2 $ and $\log(N_{H}/\ccol) = 20$. Red, green and blue lines represent models with the $\log U = -2 ,\, 0\, {\rm and}\, 1$, respectively. The offset in intensity between spectra has been inserted only for visual clarity.}
	\label{fig:spectra}
\end{figure*}

\section{Overview of emission lines diagnostics}\label{sec:overview}
In this section we briefly discuss some popular emission line diagnostics used to estimate specific physical parameters of the ISM of galaxies. For an extensive review on emission lines we refer the interested reader to \citet{Stasinska2007}.
	
Let us first consider the classical case of density indicators. This is often done by using two similar energy transitions (but different transition probabilities) of a given ion \citep{Osterbrock1989}. Ions (transitions) typically used are the [OII] ($\lambda$3726, 3729) or [SII] ($\lambda$6716, 6731). In both cases the transitions are excited from the ground level to two slightly different upper levels; they correspond to different critical densities. The intensity ratio of the two lines is sensitive to gas density.
	
One of the most popular indicators for the metallicity of the gas is the R$_{23}$ parameter proposed by \citet{Pagel1979}. This is defined as follows:
	
\begin{equation}
R_{23} = \frac{\text{I([OII]}\lambda 3727) + \text{I([OIII]}\lambda 4959) + \text{I([OIII]}\lambda 5007)}{\text{I(H}\beta)}\,,
\label{eq:r_23}
\end{equation}
where \emph{I} denotes the emission line intensity. A problem is that the indicator is non-monotonic, i.e. for a given value of R$_{23}$, two different metallicity values are possible solutions. To break this degeneracy, additional diagnostics have been proposed. However most of these methods rely on the [NII]$\lambda$6584 line \citep{Pettini2004}, that is usually very weak and tends to become difficult to measure especially at low metallicities, i.e. [NII]$\lambda$6584 / H$\alpha$ $<$ 0.1 at Z $\lesssim$ 0.6 Z$_{\odot}$.

Another alternative to the metallicity calibration is the S$_{23}$ parameter, introduced by \citet{Dìaz2000} This is based on the use of the sulphur abundance parameter S$_{23}$ \citep{Vìlchez1996}:
	
\begin{equation}
S_{23} = \frac{\text{I([SII]}\lambda \lambda 6717,6731) + \text{I([SIII]}\lambda \lambda 9069,9532)}{\text{I(H}\beta)}\,.
\end{equation}
	
With the recent availability of large data sets with known metallicity, empirical calibrations for metallicity diagnostics over a relatively large range of values have been proposed \citep{Nagao2006,Maiolino2008,Nagao2011}.
	
Other useful indicators, along with a critical analysis of different techniques, can be found in \citet{Kewley2008}. These authors suggest the use of calibrators as [NII]$\lambda$6584 / H$\alpha$ \citep{Kewley2002} or [NII]$\lambda$6584 / [OII] ($\lambda$3726,3729) \citep{Pettini2004}, since these two methods give low residual discrepancies in the estimation of metallicities thus providing also new \quotes{revised} formulae for these calibrators.
	
All the previous works typically use rest-frame optical diagnostics. Therefore they are considerably affected by dust extinction, and additional assumptions must be made in order to apply a differential correction to line intensities. As such, optical and near-InfraRed (IR) indicators can be considered fully reliable only in the outer and less extinguished region of galaxies. However, their properties may deviate substantially from the conditions found in the inner, more active, and more obscured regions.
	
To cure this problem, \citet{Nagao2011} proposed a metallicity diagnostic based on IR fine structure emission lines. They find that a reliable indicator can be the ratio:
	
\begin{equation}
R = \frac{\text{I([OIII]}\lambda 51.80 \mu m) + \text{I([OIII]}\lambda 88.33 \mu m)}{\text{I([NIII]}\lambda 57.21 \mu m)}\,.
\end{equation}
	
Both for the optical and IR lines, the accuracy of the metallicity determination is still poor because of the indicator dependencies on parameters other than metallicity, such as density and ionization parameter \citep{Nagao2012}.

Star Formation Rate (SFR) line diagnostics in the optical and near-IR wavelength range ($\lambda \sim$ 3,000 - 25,000 \angstrom) are derived from hydrogen recombination lines (e.g. H$\alpha$, H$\beta$, P$\alpha$) and forbidden lines  (e.g. [OII] and [OIII]). These lines effectively trace the ionizing radiation field in the system. For a stellar population with constant SFR over the past 100 Myr, characterized by a Kroupa Initial Mass Function (IMF, \citet{Kroupa2001}, slope -1.3 in the range 0.1 - 0.5 M$_{\odot}$ and slope -2.3 in the range 0.5 - 120 M$_{\odot}$), \citet{Calzetti2008} reports:
	
\begin{equation}
\text{SFR}/(\msun\, {\rm yr}^{-1}) = 5.3\, \text{L(H}\alpha)/(10^{42} \text{ erg s}^{-1})\,,
\label{eq:calzetti_sfr}
\end{equation}
where L(H$\alpha$) has been corrected for underlying stellar absorption and interstellar dust attenuation, barring variations of $\pm 20\%$ for younger/older ages and different metallicities. Note that, with respect to the relation provided by \citet{Kennicutt1998}:
	
\begin{equation}
\text{SFR}/(\msun\, {\rm yr}^{-1}) = 7.9\, \text{L(H}\alpha)/(10^{42} \text{ erg s}^{-1})
\label{eq:kennicutt_sfr}
\end{equation}
there is a difference of about 50\%. This is mainly due to different assumptions for the IMF and age of the stellar populations.
	
Far-IR (FIR) lines ratios can be also used as probes of several nebular properties \citep{Rubin1994,Spinoglio2015}. Useful prescriptions to estimate the SFR are those reported in \citet{DeLooze2014}. These authors used FIR fine-structure lines (158$\mu$m [CII], 63$\mu$m [OI] and 88$\mu$m [OIII]) and the relation reported in \citet{Vallini2015} to connect the luminosity of the [CII] line with the SFR and the metallicity of the gas \citep[see also][]{Pallottini2015}. 
	
Finally, using a combination of stellar population synthesis and photoionization models, \citet{Kewley2002} developed a set of ionization parameter and metallicity diagnostics based on the strongest optical emission lines from HII regions.

\section{Library of synthetic spectra}
We use the photoionization code \textlcsc{CLOUDY v13.03} \citep{Cloudy} to create a large library of synthetic spectra. The inputs are grids of ISM physical properties: total gas density ($n$), column density ($N_{H}$), metallicity ($Z$) and ionization parameter ($U$) chosen within the ranges shown in Table \ref{table:grid}, The ranges for $n$ and $N_{H}$ are chosen to cover the different ISM phases. 
	
The density spans from values typical of the Hot Ionized Medium (HIM, $n\sim 10^{-3} {\rm cm}^{-3}$) to those of dense molecular cores ($n \sim 10^{5} {\rm cm}^{-3}$); the column density goes from the one expected in HII regions ($N_{H} \sim 10^{17} {\rm cm}^{-2}$) to dense clouds ($N_{H} \sim 10^{21} \div 10^{23} {\rm cm} ^{-2}$). Metallicity varies from extremely metal poor cases ($Z\sim 10^{-3} \zsun$) to slightly super-solar values ($Z \sim 3 Z_{\odot}$).

\begin{table}
	\caption{Range of ISM physical parameters explored in this work.}
	\centering
	\begin{tabular}{c  c c}
		\hline\hline
		Parameter & minimum & maximum\\
		\hline
		$\log(Z / \zsun)$ & -3.0 & 0.5\\
		$\log(n / {\rm cm}^{-3})$ & -3.0 & 5.0\\
		$\log(U)$ & -4.0 & 3.0\\
		$\log(N_{H} / {\rm cm}^{-2}) $ & 17.0 & 23.0\\
		\hline
		\hline
	\end{tabular}
	\label{table:grid}
\end{table}

\subsection{Definitions}
The ionization parameter is the number of ionizing photons per hydrogen atoms, and we adopt the following definition \citep[e.g.][]{Yeh2012}:
	
\begin{equation}
U = \frac{1}{4\pi R_{S}^{2} n c} \int_{\nu_{e}}^\infty \frac{L_{\nu}}{h \nu} d\nu = \frac{Q(H)}{4\pi R_{S}^{2} n c}\,,
\label{eq:ionization_parameter}
\end{equation}
where $Q(H)$ is the ionizing photon flux, $c$ the speed of light and $R_{S}$ is the Str\"{o}mgren radius \citep{Stromgren1939}:
	
\begin{equation}
R_{S}^{3} = \frac{3 Q(H)}{4\pi n^{2} \alpha(T)}\,,
\label{eq:stromgren_radius}
\end{equation}
where $\alpha$ is the temperature ($T$) dependent recombination rate. By combining eq.s \ref{eq:ionization_parameter} and \ref{eq:stromgren_radius}, we obtain:

\begin{equation}
U = \frac{1}{c} \sqrt[3]{\frac{Q(H) n \alpha^2}{36\pi}}\,.
\end{equation}

The minimum ionization parameter inside a HII region generated by a single early spectral type star can be estimated as follows. If we assume a density $n\sim 100\,{\rm cm}^{-3}$, a recombination rate $\alpha \sim 2.6 \cdot 10^{-13}{\rm cm}^{3} {\rm s}^{-1}$, and consider the $Q(H)$ from a star with $Z_{\star}\simeq \zsun$ and mass $M_{\star} = 10\,\msun$, we get $\log U_{min} \sim -4.0$ \citep{Schaerer2002}. Hence this justifies the minimum value for $\log U$ in Table \ref{table:grid}.
	
\begin{figure}
	\centering
	\includegraphics[width=0.85\linewidth]{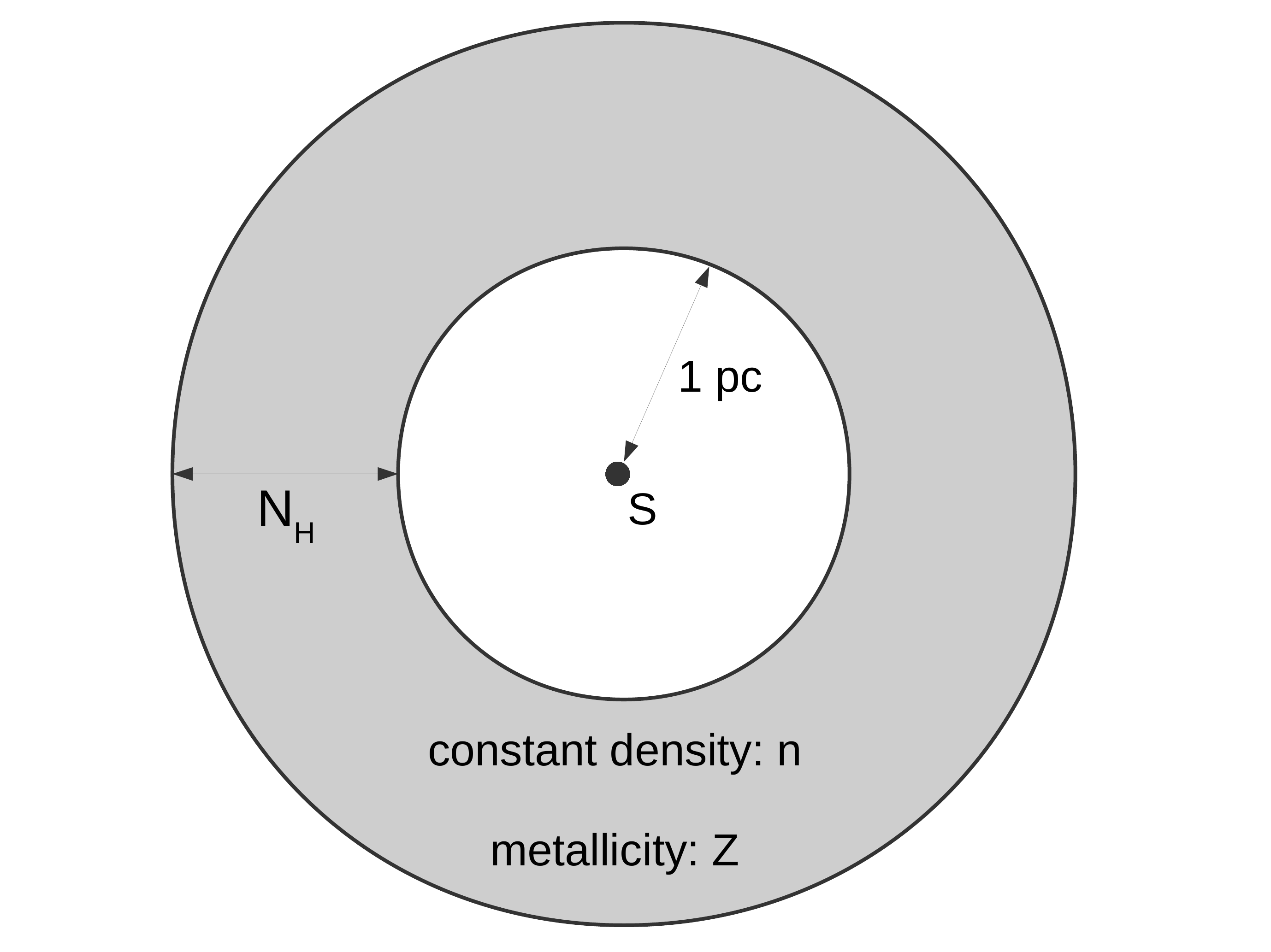}
	\caption{Geometry adopted to build our library. A central source S (with a given ionization parameter $U$) illuminates the inner part of the cloud located at 1 parsec from the center. The density ($n$) and metallicity ($Z$) of the cloud are assumed to be constant. The outer radius of the cloud is the distance at which our column density ($N_{H}$) reaches the required value for the model under consideration.}
	\label{fig:toy_model}
\end{figure}

For our \textlcsc{CLOUDY} calculations we consider a spherical geometry and static conditions. A central ionizing source illuminates the inner part of the cloud situated at distance 1 pc from the center (see Fig. \ref{fig:toy_model}). The outer part of the cloud defines the end of the calculation. For each model ($n={\rm cost}$), the outer cloud distance is set in order to reach the required value for the column density (see Table \ref{table:grid}). We have therefore removed the default stopping criterion based on a lower limit on the gas kinetic temperature (4000 K).

Metal abundances for all calculations are assumed to be solar \citep{Grevesse2010}. For dust we consider contribution from graphite and silicate components that reproduce the observed overall extinction properties of the MW ISM: R$_{V}$ = A$_{V}$ / E(B - V) = 3.1. The grain size distribution is described by a power-law distribution \citep{Mathis1977} resolved by default in \textlcsc{CLOUDY} into ten size bins. We did not include the contribution from PAHs in the library used in this work. Observations for local galaxies with lower metallicities (e.g. IZw18 and SBS0335-052), show in fact a suppressed PAH emission \citep{Hunt2010,Wu2009,Wu2007} and these are the prototypes for the high redshift galaxies (z $\sim$ 6) we are interested in (see Section \ref{sec:input_features}). The temperature within the cloud is computed by \textlcsc{CLOUDY} from the balance between heating and cooling.

\subsection{Stellar Model Spectra}
The flux illuminating the cloud defines the ionization parameter $U$ (see eq. \ref{eq:ionization_parameter}) that can be computed by using the \textlcsc{STARBURST99} code \citep{Leitherer,Smith2002,Leitherer2014}. This code computes theoretical spectral energy distributions (SED) combining a stellar atmosphere models with grids of stellar evolutionary tracks that account for different metallicities, star formation histories, IMF, and age of stellar populations. 
	
We adopted a Salpeter IMF \citep{Salpeter} in the $0.1 - 120 \msun$ mass range and a constant star formation rate of 1 $\msun\,{\rm yr}^{-1}$. The metallicities of the input spectra considered are $Z =\, $0.001, 0.008, 0.020 ($\zsun$) and 0.040. We calculated models using the set of evolutionary tracks produced by the Geneva group \citep{Schaller1992}. We use the \quotesing{Lejeune/Schmutz} option that adopts the extended model atmospheres of \citet{Schmutz1992} in the case of stars with strong winds, and the plane-parallel atmospheric grid of \citet{Lejeune1997} otherwise. We fix the stellar population cluster age to 10 Myr. Note that the shape of the EUV spectrum ($100$\angstrom$ <\lambda<1000$ \angstrom) does not change for older stellar populations \citep{Kewley2001}. 
	
STARBUST99 spectra obtained with these prescriptions are given as input to the \textlcsc{CLOUDY} code. Different examples of the emerging spectra are reported in Fig. \ref{fig:spectra}, for $Z=0.005~\zsun$ (upper panel) and $Z=0.5~\zsun$ (lower panel) and different values of the ionization parameter ($\log U=-2,0,1$), at fixed $n=10^{2}\cc$ and $N_H=10^{20}~\rm cm^{-2}$. The spectrum includes the stellar and dust continuum with superimposed the emission lines of hydrogen, helium and the major elements commonly found in the ISM of galaxies. The figure also shows that the IR/FIR peak due to the dust continuum emission is shifting towards larger wavelengths (i.e. colder dust) at decreasing ionization parameter.
	
Each \textlcsc{CLOUDY} output spectrum is then labeled with its corresponding parameters and stored in the library. Then, we implement and train a Supervised ML (SML) algorithm, described in the following section, that allows us to recover the parameters associated with any given input spectrum.

\section{Machine Learning Methods}
In this section, we describe the SML approach used in this work, and we briefly review its main algorithms, namely Decision Trees and AdaBoost. The main idea of SML is that an observable quantity (i.e. a spectrum) is a set of $x$'s (e.g. spectral lines) that we relate to a set of $y$'s (i.e. the four physical properties $n,\, Z,\, N_{H},\,{\rm and}\,U$). The task is to use a training set in order to find  an algorithm $f(x)$ such that for future $x$ in a test set, it will be a good predictor of $y$.

\begin{figure}
	\centering
	\includegraphics[width=0.8\linewidth]{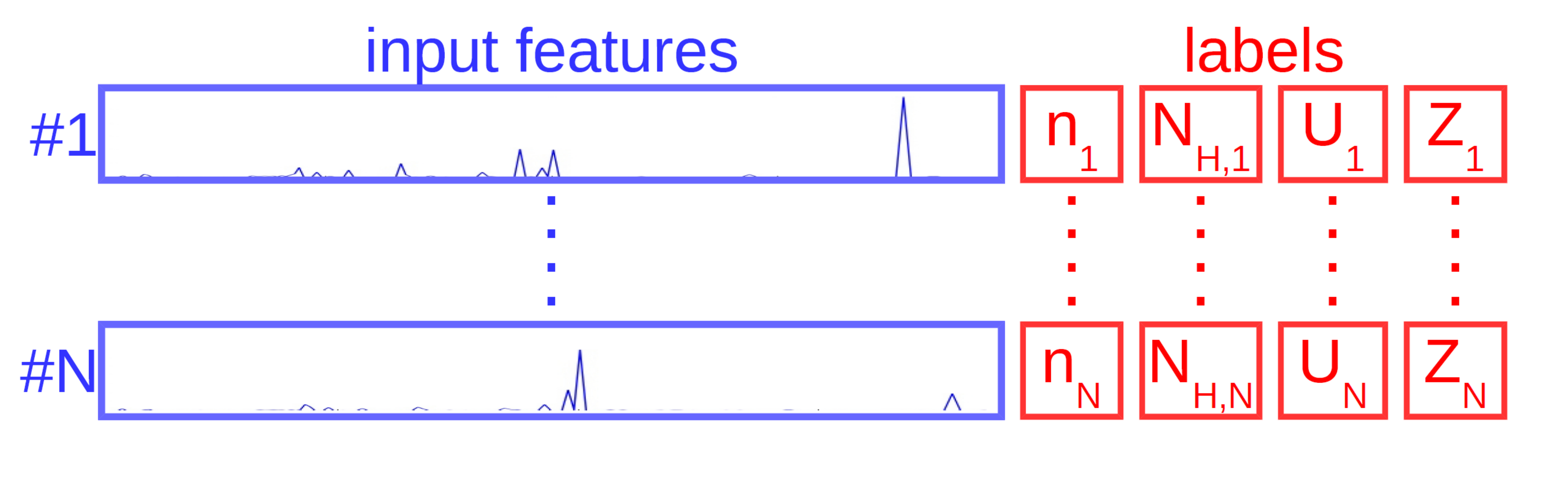}
	\caption{Visual representation of the training dataset for our Supervised Machine Learning (SML) method. The dataset is composed by a set of N spectra ($3\times 10^4$ in our case); each of them in turn consists of a set of input features (blue). To each spectrum is assigned a label (red) that represents the physical property that the model, once trained, tries to predict.}
	\label{fig:input_array}
\end{figure}

The SML method tries to infer the physical properties of a given input from labeled data. The training dataset consists of a set of input features (i.e. an input vector) and a set of labels (i.e. the desired output values) for each example (see Fig. \ref{fig:input_array}). The SML algorithm analyzes the training dataset and produces a model that ideally should give as output the same labels required for the training process. This model can be used at this point to predict the physical values of new input examples.

For the analysis performed in this work, we used the scikit-learn \citep{scikit-learn} Python package\footnote{http://scikit-learn.org/}.

\subsection{Decision Trees based methods}
Decision Trees \citep{Breiman1984,scikit-learn} are supervised algorithms commonly used in ML, whose main purpose is to create a model in the form of a tree structure that is able to predict the value of a target variable starting from a vector of several input variables. Decision trees algorithms find final decision boundaries automatically based on the data. They are valid for a very large range of applications and are also extremely fast. 

The core algorithm for building decision trees is the ID3 (Iterative Dichotomizer 3) developed by \citet{Quinlan1986}. This algorithm creates a tree and determines for each node the feature yielding the largest information gain for targets. C4.5 \citep{Quinlan1993} is the successor to ID3 that has overcome some restrictions of the previous algorithm. The new main features of C.4 are: (i) it accepts both categorical and numerical variables, (ii) it better handles over-fitting by applying \quotes{pruning} techniques \citep{Breiman1984,Mingers1989,Mehta1995}. The scikit-learn package we used, adopts an optimized version of the CART (Classification and Regression Trees) algorithm based on C4.5. CART constructs binary trees using the input features (see Fig. \ref{fig:input_array}) and thresholds that yield the largest information gain at each node.

A decision tree recursively partitions the input features space into an increasing number of \quotes{leaves}. Each leaf represents a value for the desired output and it is chosen to minimize the mean squared error of the output labels, that in our case are the physical parameters defined in Table \ref{table:grid}.

\subsection{AdaBoost}
Ensemble learning methods \citep{Dietterich2000} are based on the assumption that many base learning algorithms, such as Decision Trees based methods, can be combined into an \quotes{ensemble system} which can achieve better predictive performances. Ensemble learners have also the advantage that they are less prone to overfit the data.

A common method to produce an ensemble of base learners is the Adaptive Boosting or AdaBoost method \citep{Freund1997,Drucker1997,Hastie}. In case of Decision Trees as the base learner, the algorithm adds trees sequentially to generate an ensemble of them. AdaBoost iteratively improves the base algorithm by accounting for the incorrectly classified examples in the training set.

First of all, equal weights are assigned to each training examples. At each step of the iteration, the base algorithm is applied to the training set and the weights of the incorrectly classified examples are increased. In each step the base learner is applied on the training set with updated weights, and after \textit{n} iterations, the final model is obtained as the weighted sum of the \textit{n} learners.

\subsection{Input Features}\label{sec:input_features}
Starting from a given synthetic SED (see Fig. \ref{fig:spectra}) obtained with the photoionization code \textlcsc{CLOUDY}, it is possible to obtain the continuum-subtracted intensities of the emission lines. The input vector for the ML algorithms (the input features in Fig. \ref{fig:input_array}) is in our case a collection of intensities associated to a discrete wavelength array and a label containing the four physical properties ($n,\, Z,\, N_{H},\,{\rm and}\,U$) of the model under consideration.

The range of wavelengths used to construct the model spans from 1216 \angstrom~(corresponding to the Ly$\alpha$ transition) to 4.0 $\mu$m. This particular choice would be equivalent to the rest-frame range in wavelengths obtained by combining the NIRSpec (0.6 - 5 $\mu$m) and MIRI (5 - 28 $\mu$m) instruments on board the James Webb Space Telescope (JWST) and observing a source located at redshift $z\sim 6$. Emission lines from warm/neutral gas are relatively weak but yet observable. For example [NI] $\lambda$5200 and [OI] $\lambda$6300,6364 have been observed in the spectra of local galaxies \citep{Cresci2015,Moustakas2006}. The strength of our method relies on the fact that the ML algorithm can learn from all the lines present in a spectrum, including the weakest one as those coming from the neutral ISM components. It will then possible to provide at least some constraints on the properties of these phases from observed spectra.

The SML code implementing all the above features will be referred from now on as \textlcsc{GAME} (GAlaxy Machine learning for Emission lines).

\section{Results}

\subsection{Predictive Accuracy}\label{sec:accuracy} 
In this section we present the results of the tests for the AdaBoost SML algorithm in terms of the predicted values of ($n,\, Z,\, N_{H},\, U$).

\begin{figure*}
	\centering
	\includegraphics[width=0.46\textwidth]{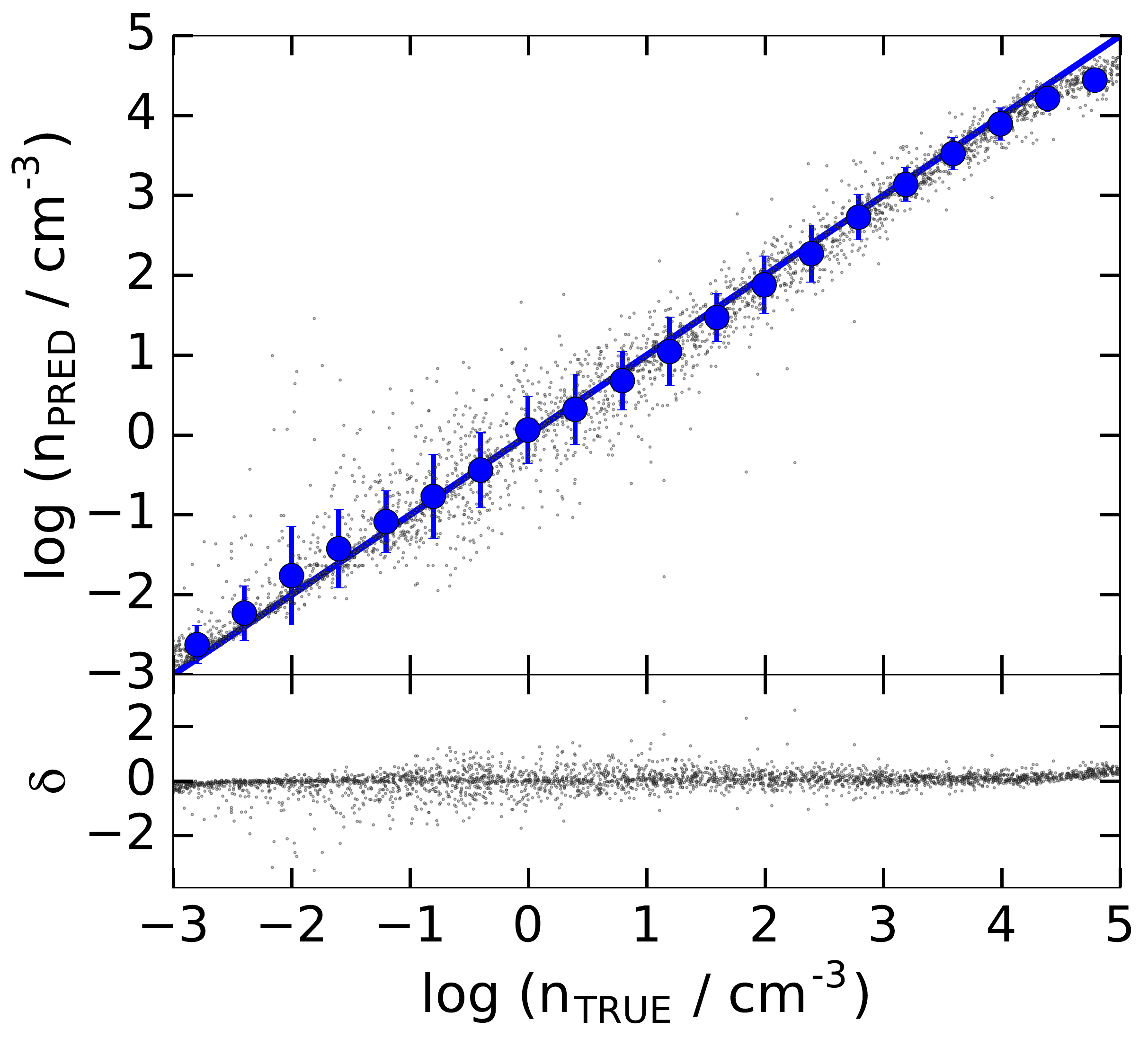}
	\includegraphics[width=0.48\textwidth]{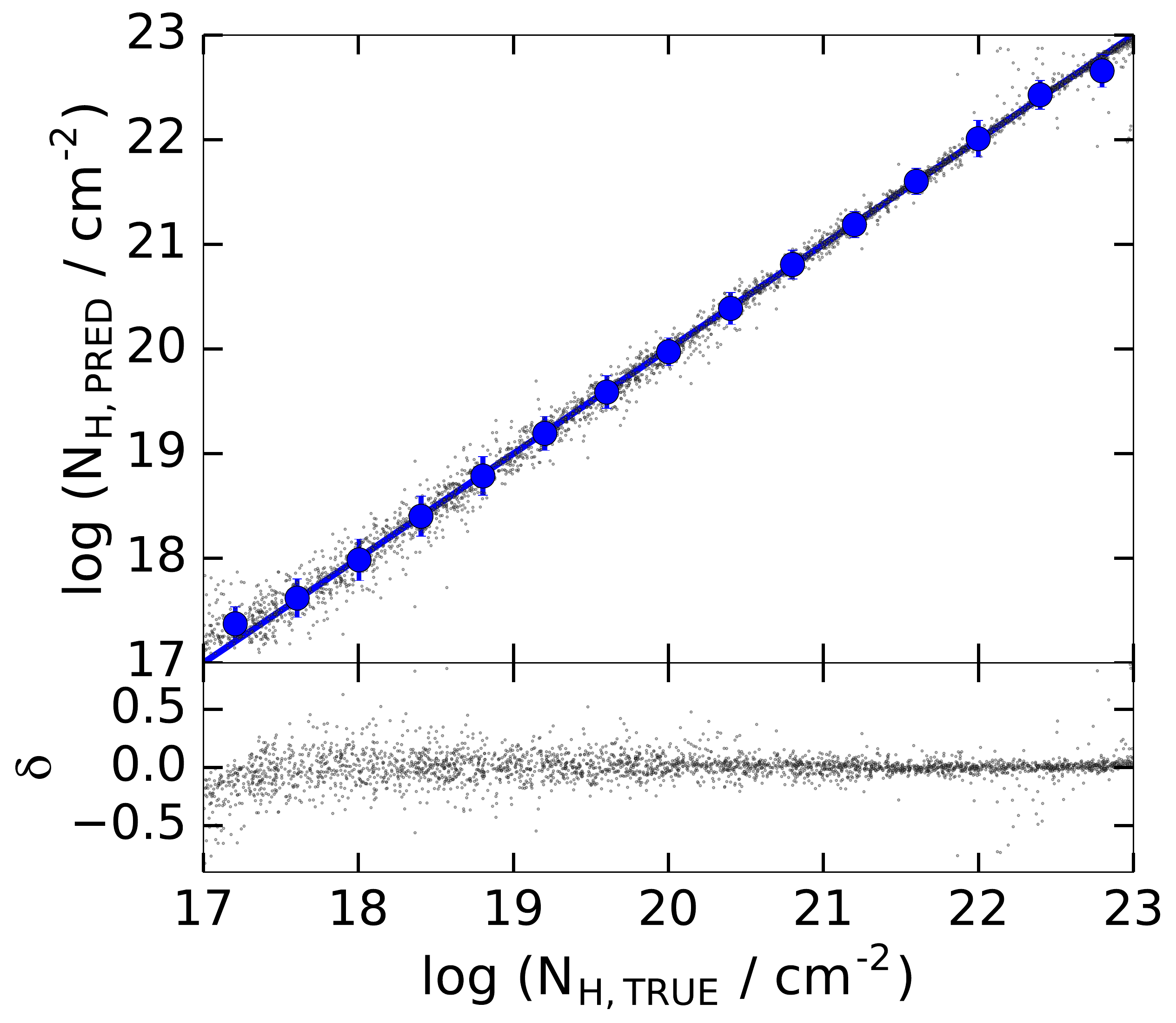}\\
    \vspace{1pt}
	\includegraphics[width=0.48\textwidth]{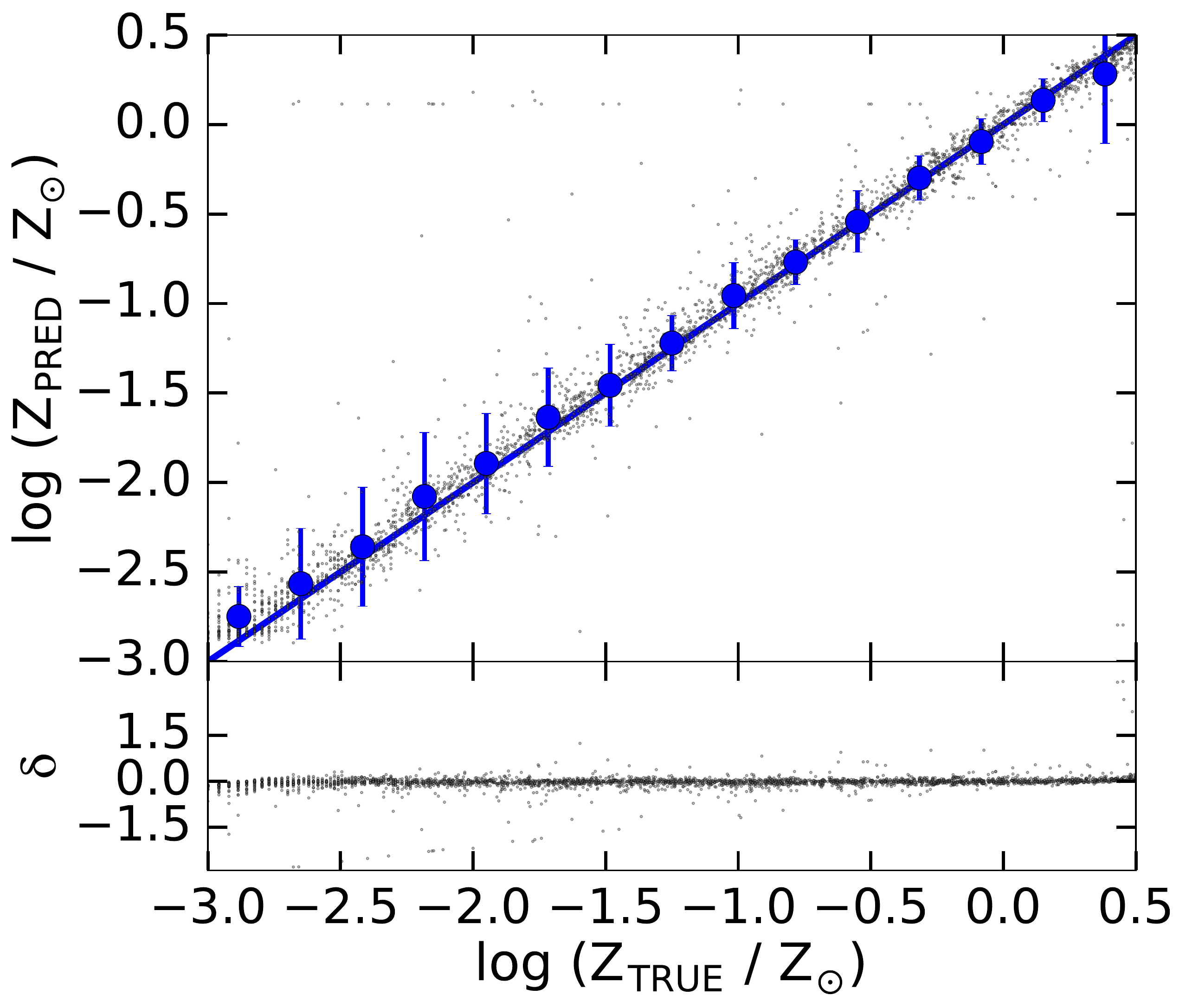}
	\includegraphics[width=0.46\textwidth]{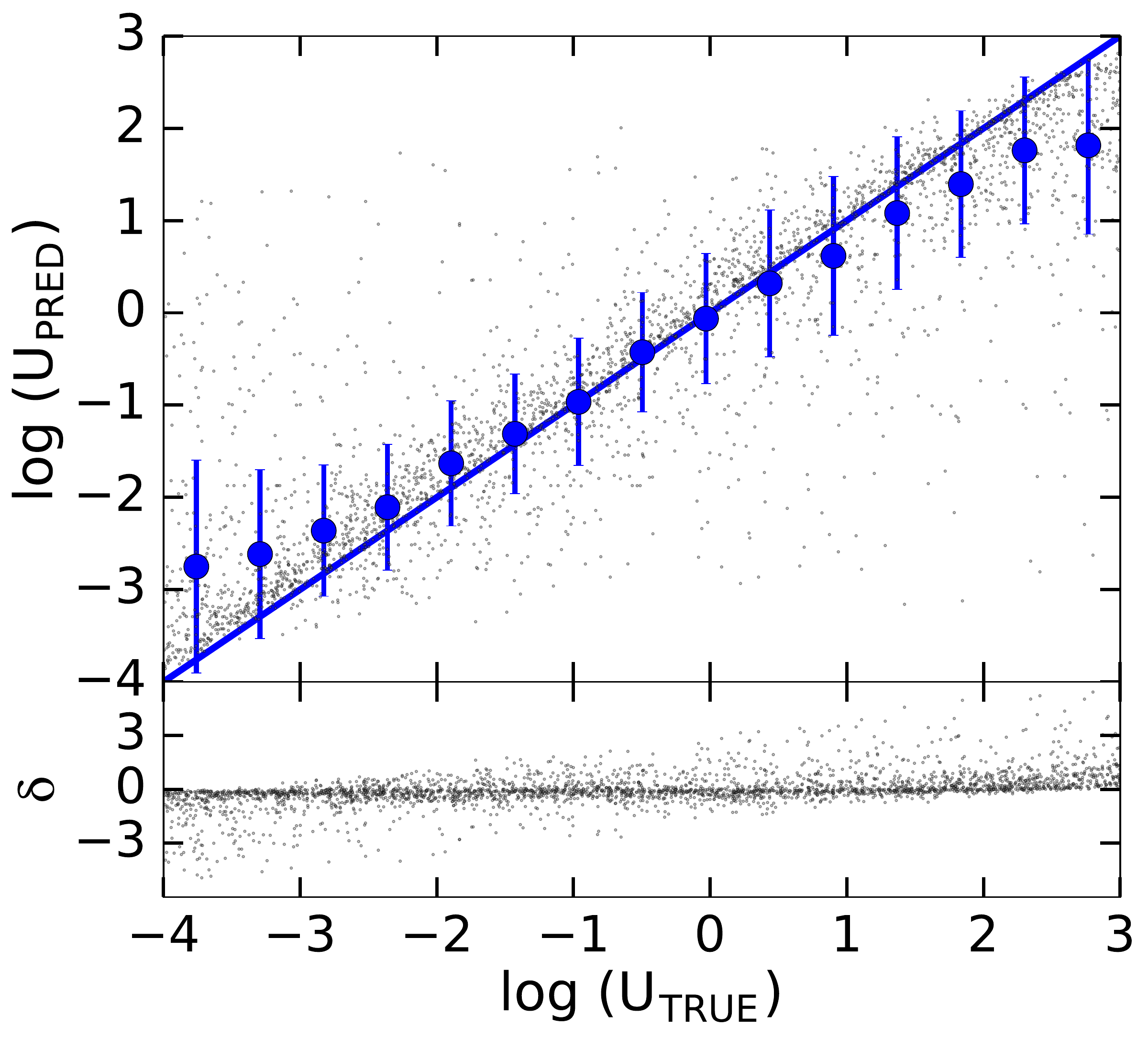}
	\caption{Scatter plots for the predicted (inferred from the model) vs true (used to construct the spectrum) physical properties (density, column density, metallicity and ionization parameter). The blue solid line is the locus of points where true = predicted. The error bars for the binned data (red) show the standard deviation of distribution. $\delta$ represents the residuals $\Delta$log (PRED - TRUE) = log (PRED / TRUE).}
	\label{fig:predicted_vs_true}
\end{figure*}

\begin{figure*}
	\centering
	\includegraphics[width=0.45\textwidth]{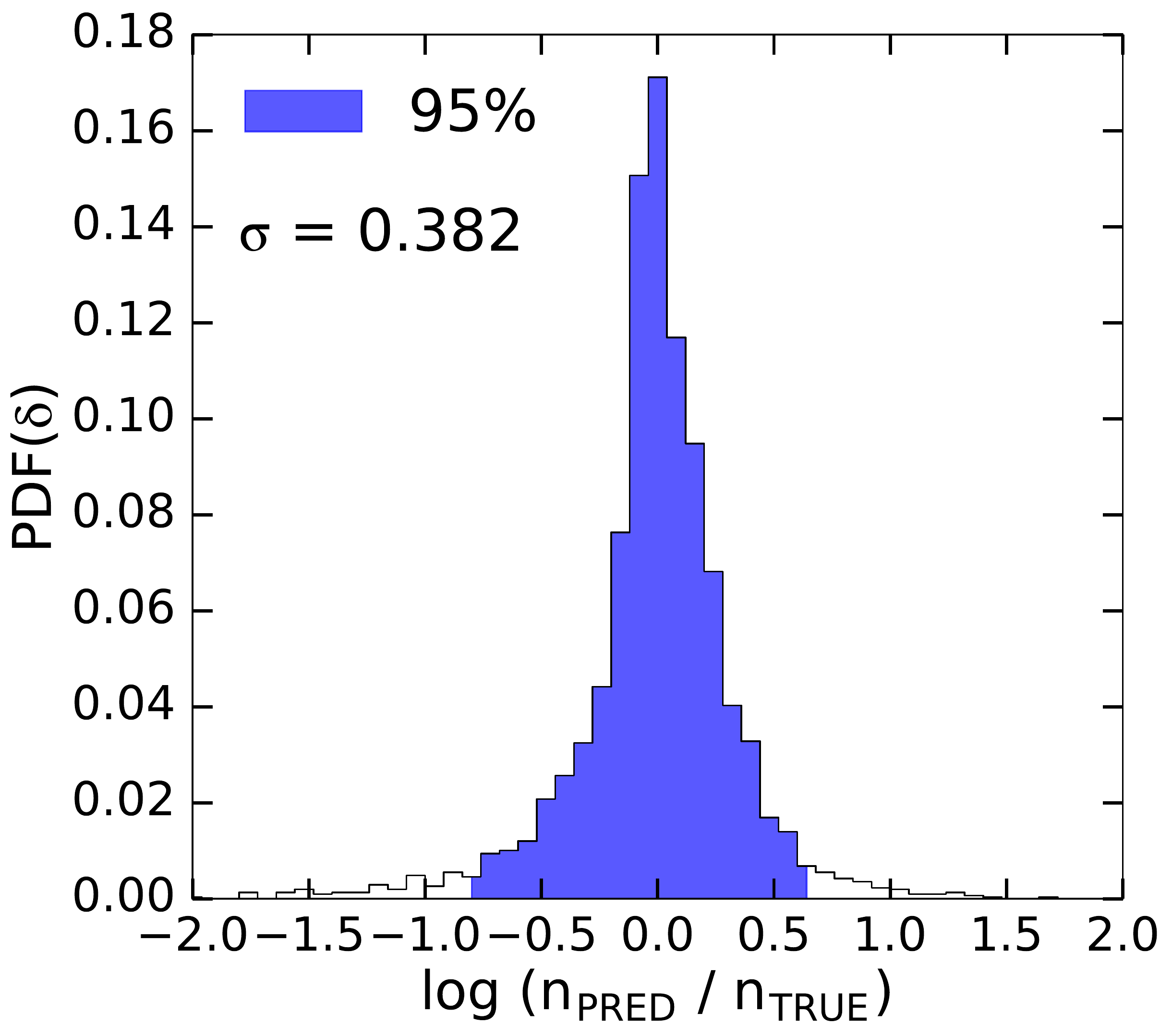}
	\includegraphics[width=0.45\textwidth]{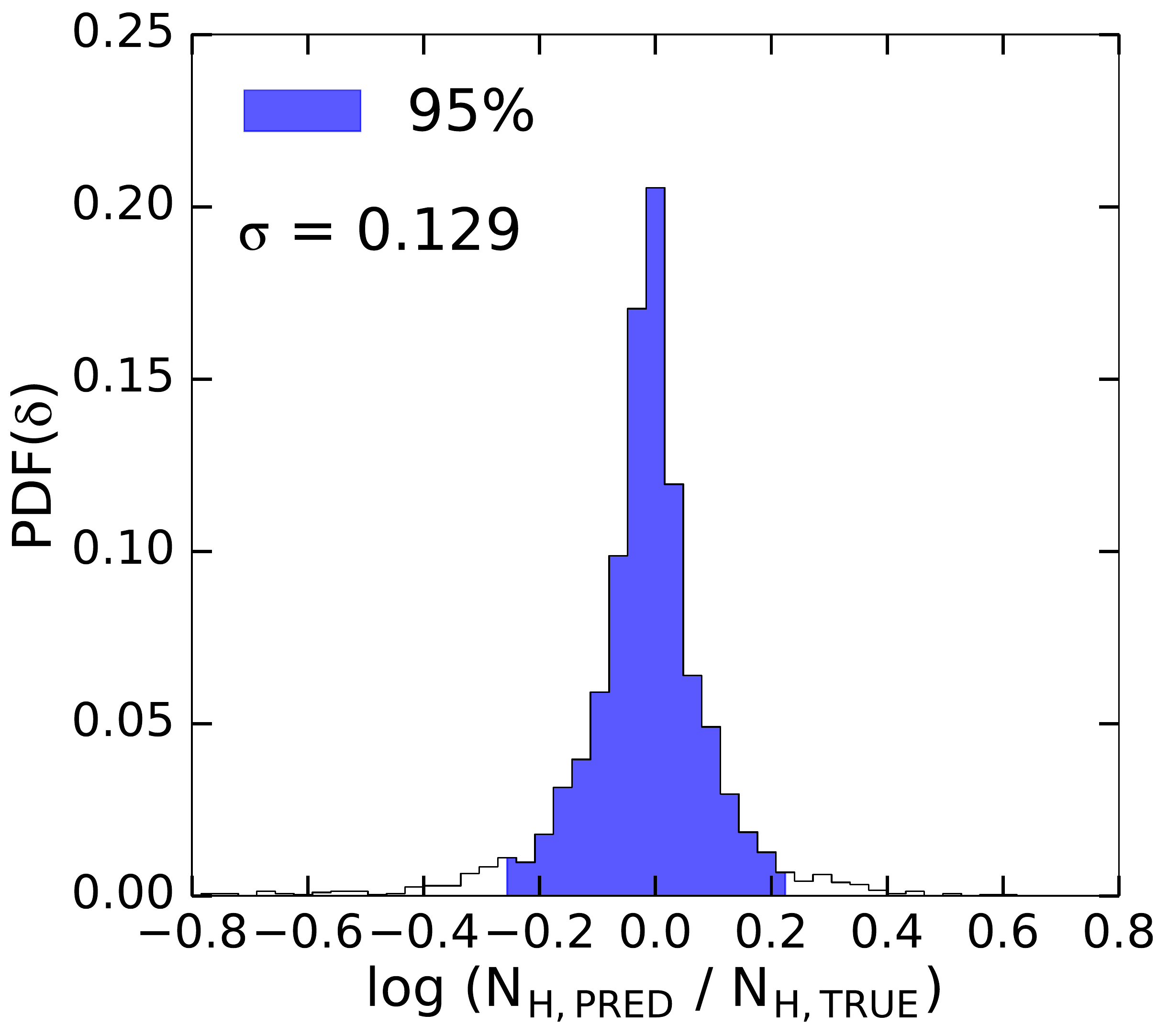}\\
	\vspace{1pt}
	\includegraphics[width=0.45\textwidth]{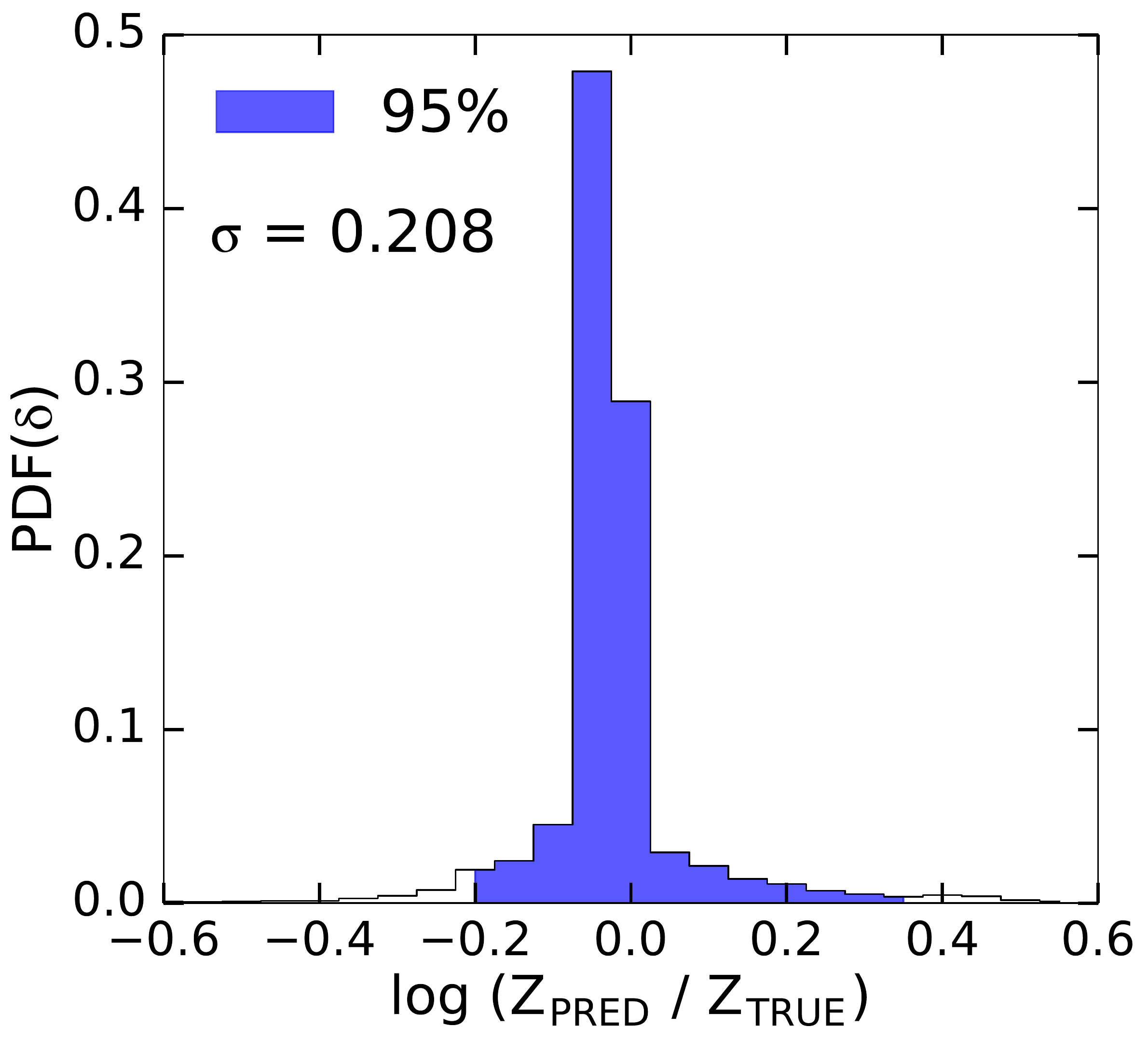}
	\includegraphics[width=0.45\textwidth]{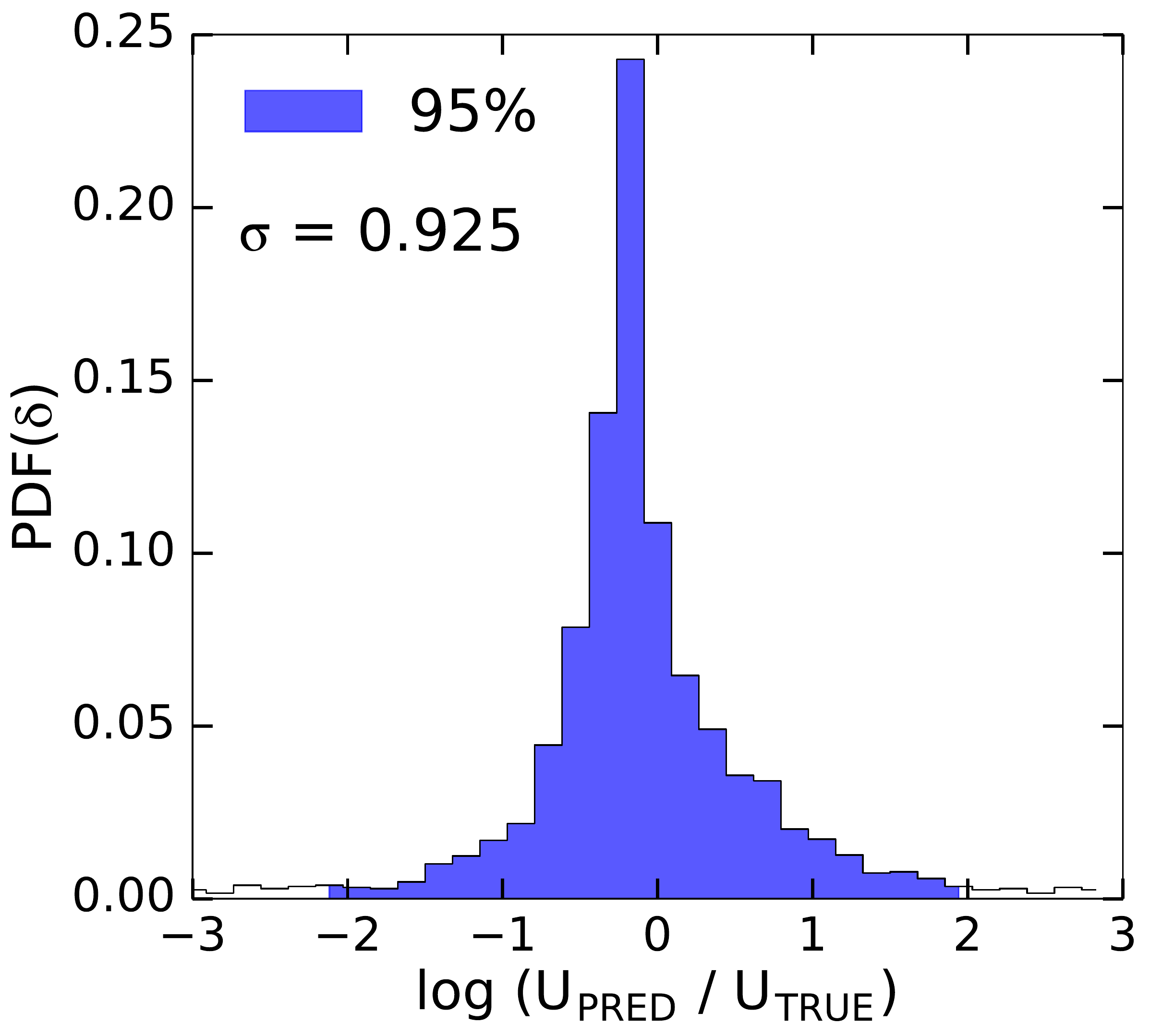}
	\caption{Probability distribution functions (PDF) of the $\delta$s reported in Fig. \ref{fig:predicted_vs_true}, i.e. the logarithm of the ratios between the predicted and true physical properties. $\sigma$ represents the standard deviation of the distributions. 95\% of the models reside within the shaded blue regions.}
	\label{fig:predicted_vs_true_hist}
\end{figure*}

\begin{figure*}
	\centering
	\includegraphics[width=0.46\textwidth]{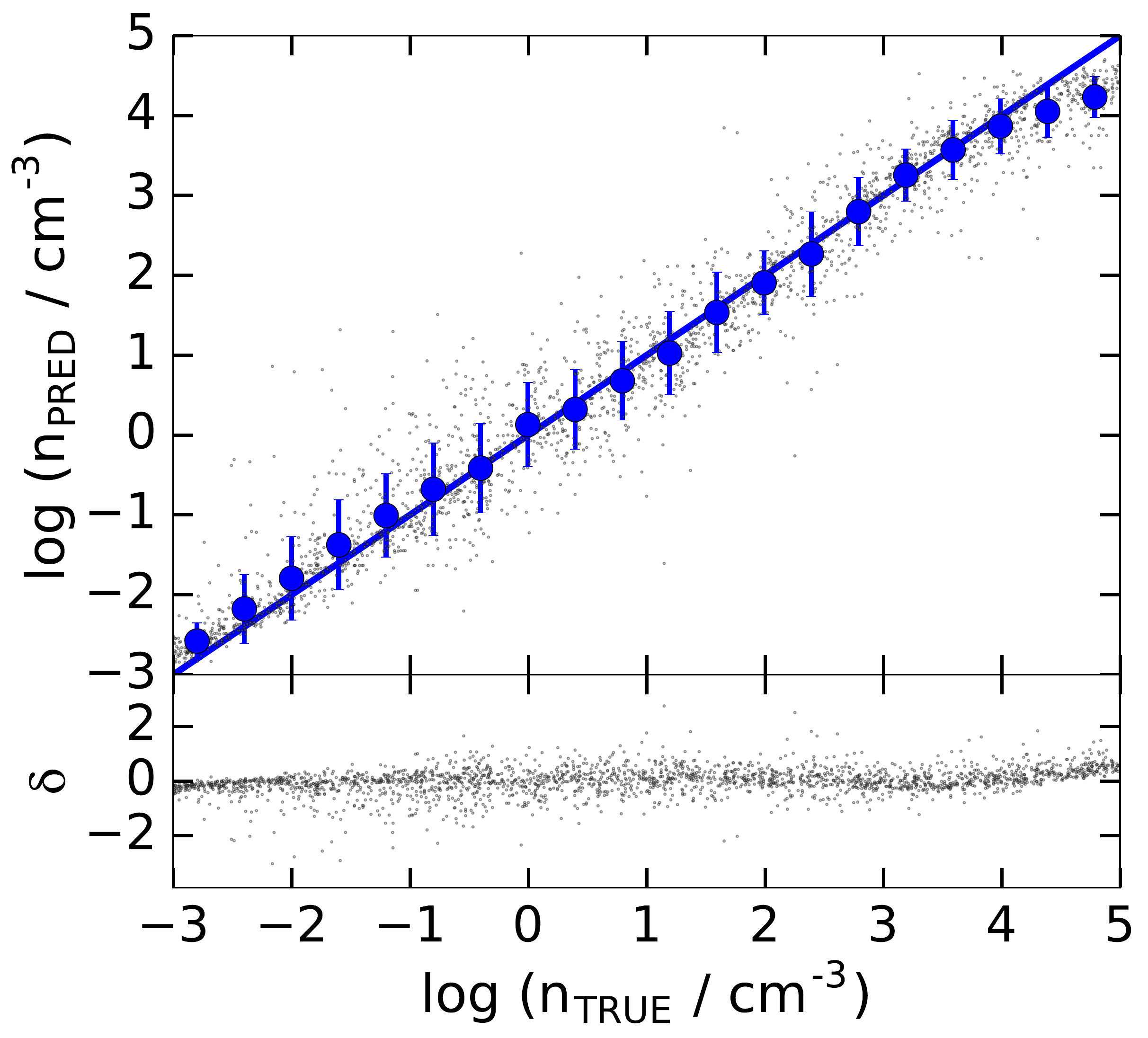}
	\includegraphics[width=0.48\textwidth]{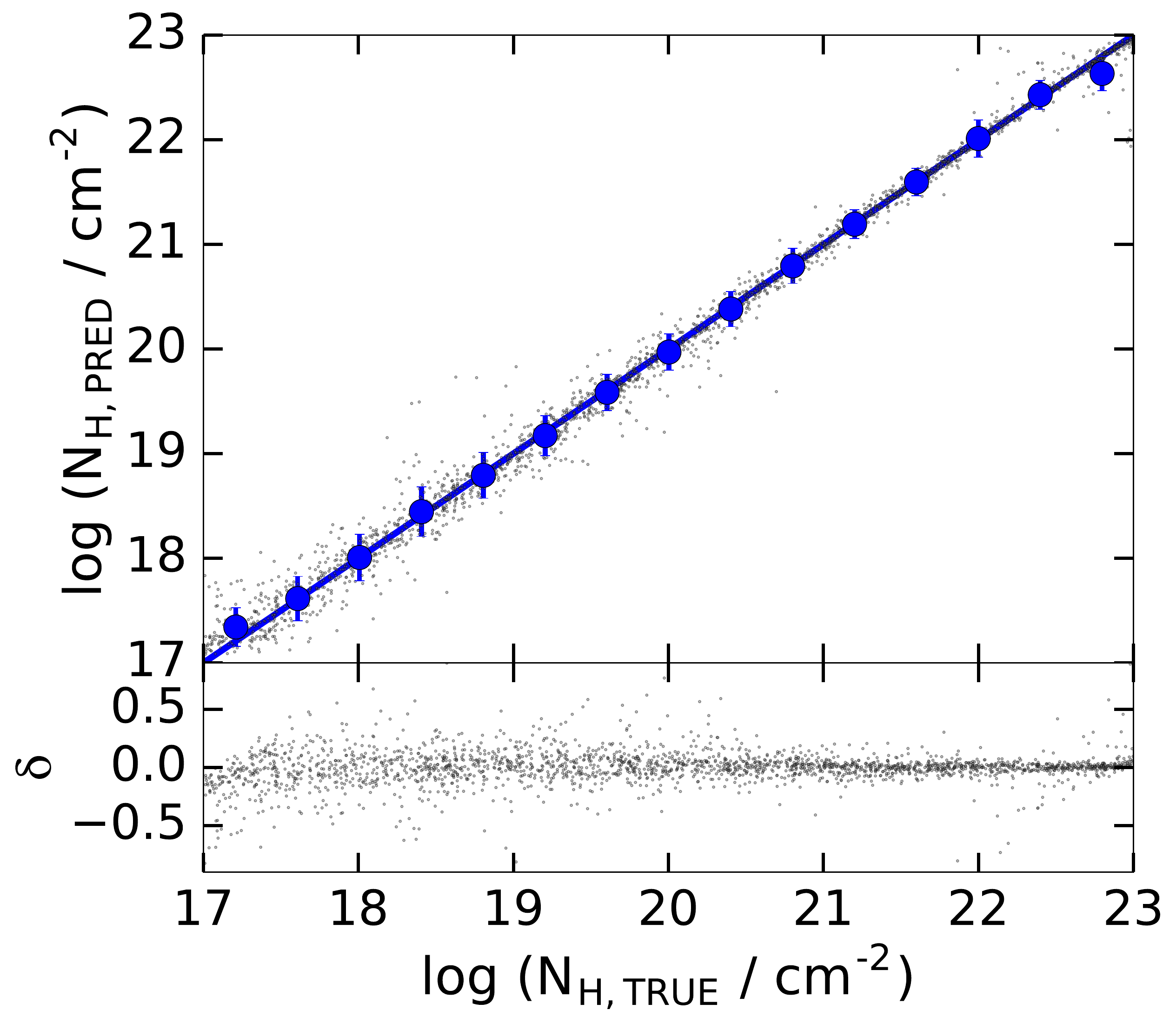}
	\includegraphics[width=0.48\textwidth]{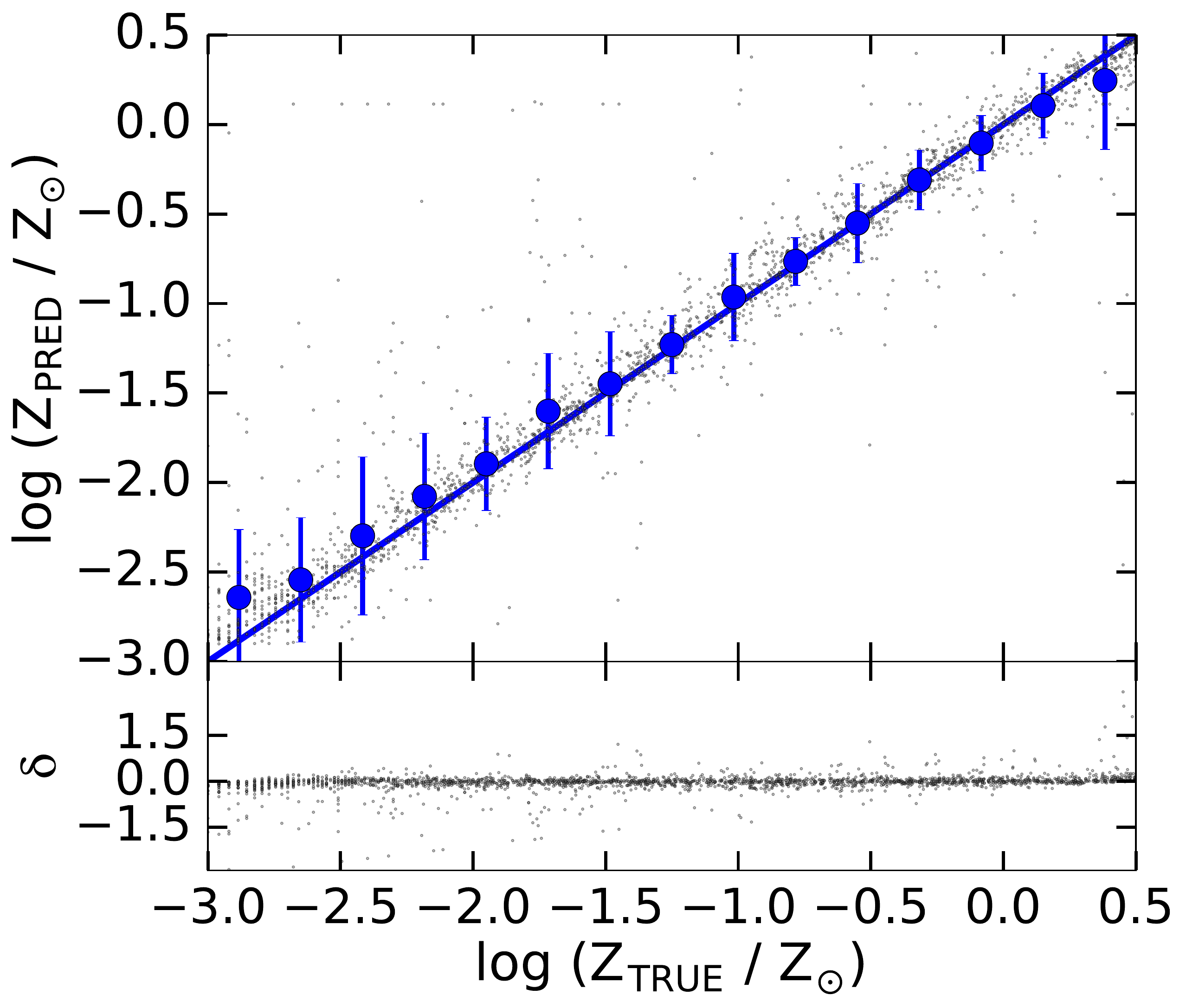}
	\includegraphics[width=0.46\textwidth]{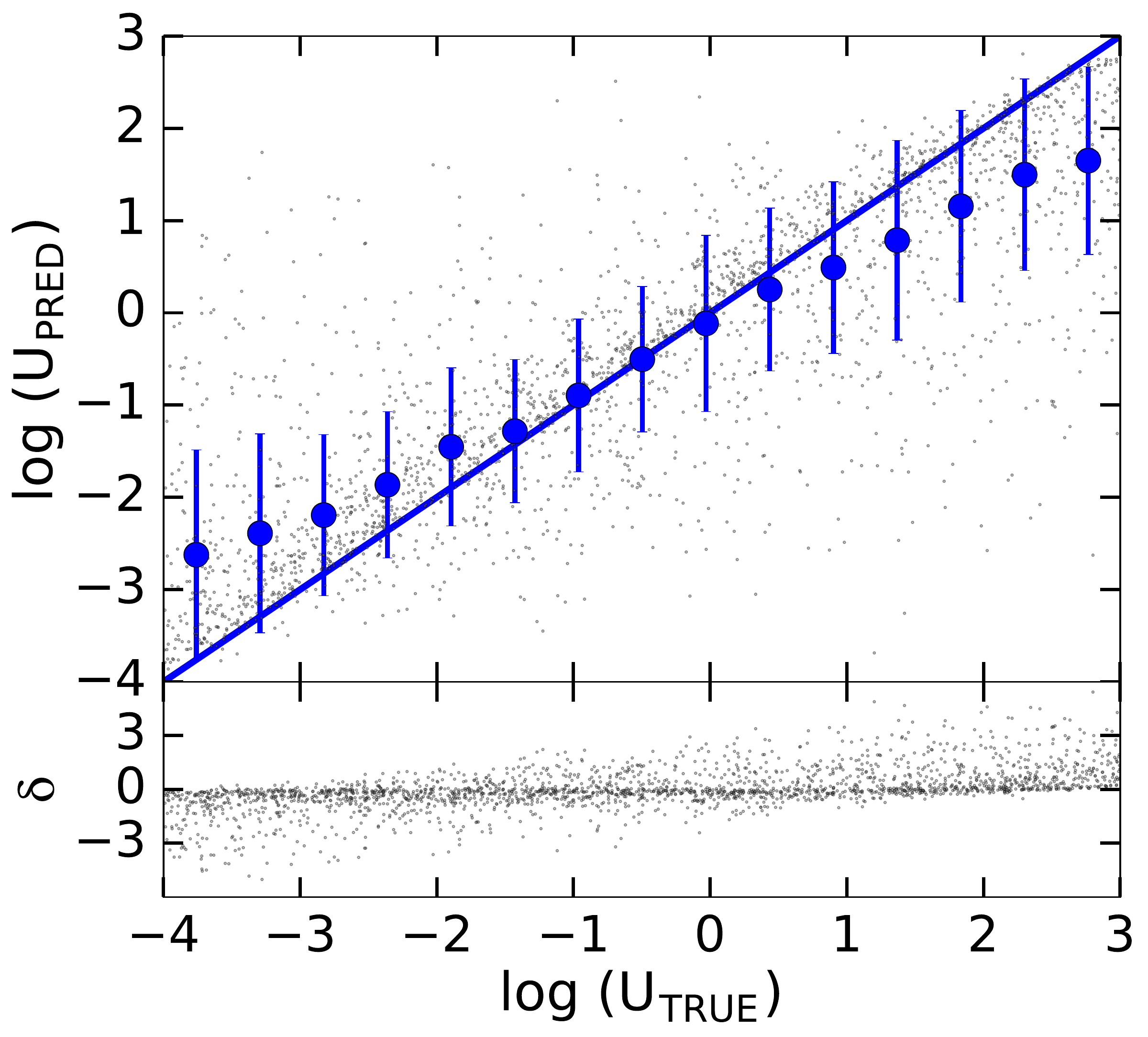}
	\caption{As in Fig. \ref{fig:predicted_vs_true}, but considering a threshold f(H$\alpha$) = 1/50 (see text for the details).}
	\label{fig:predicted_vs_true_sigma}
\end{figure*}

\begin{figure*}
	\centering
	\includegraphics[width=0.45\textwidth]{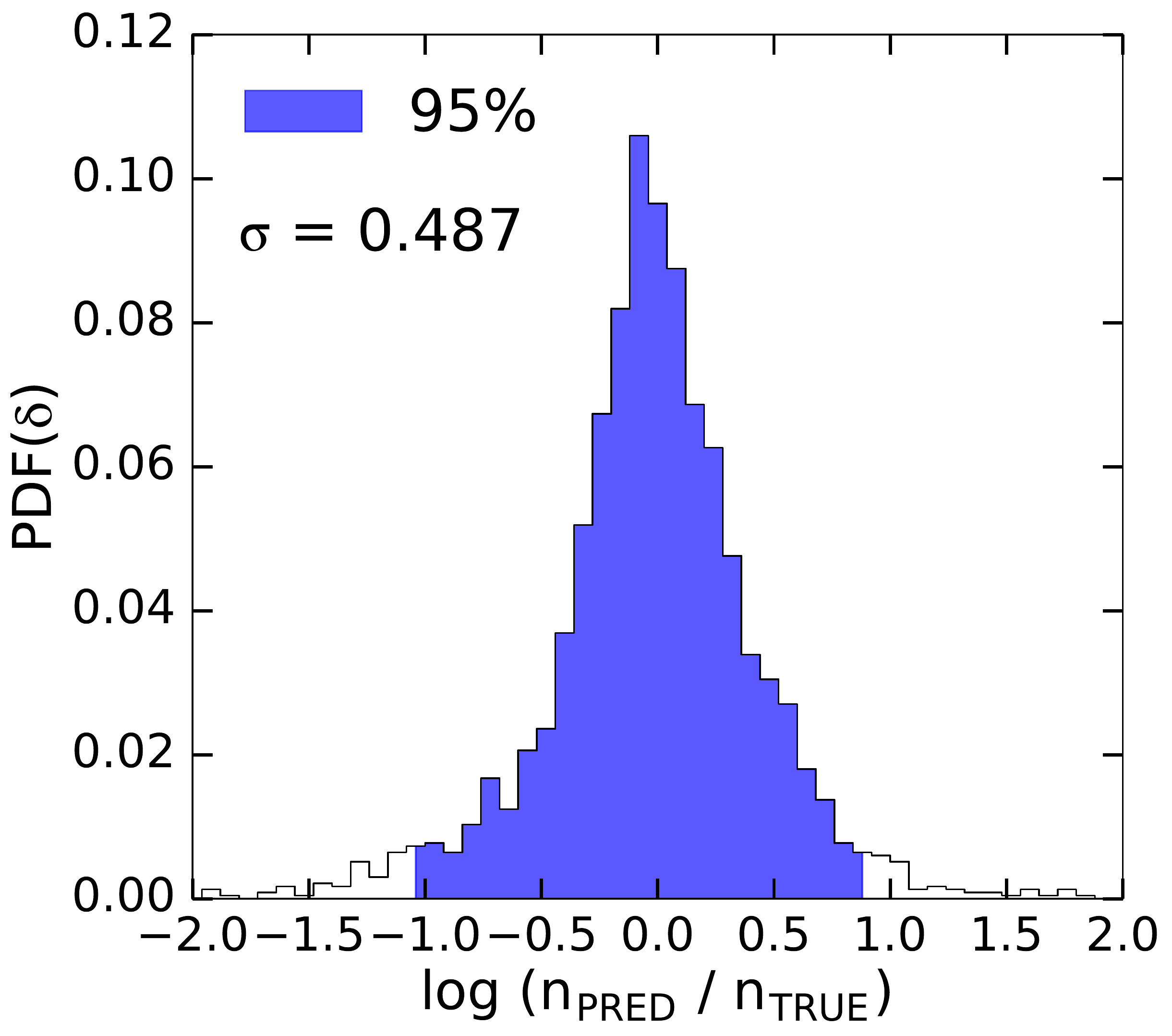}
	\includegraphics[width=0.45\textwidth]{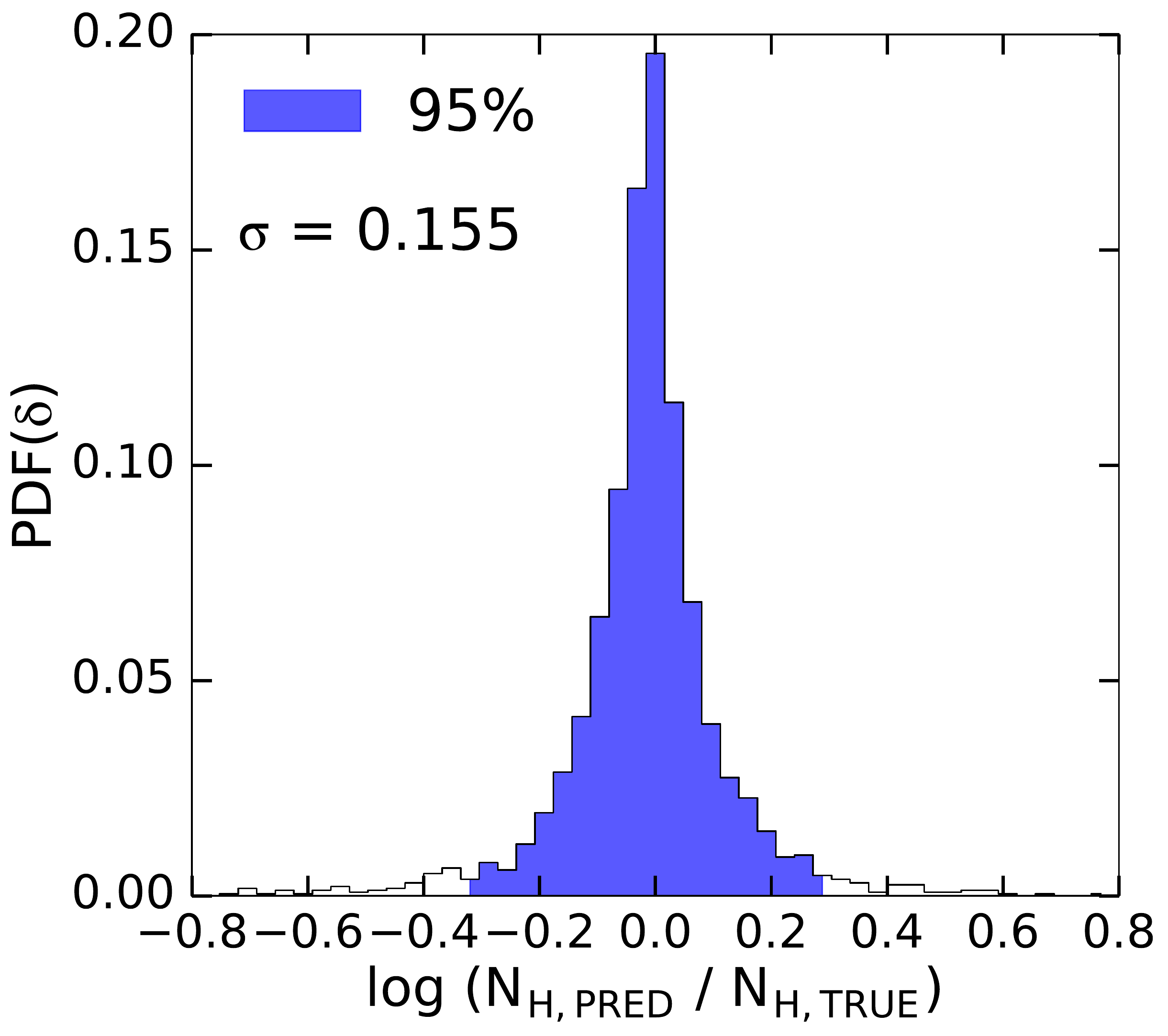}
	\includegraphics[width=0.45\textwidth]{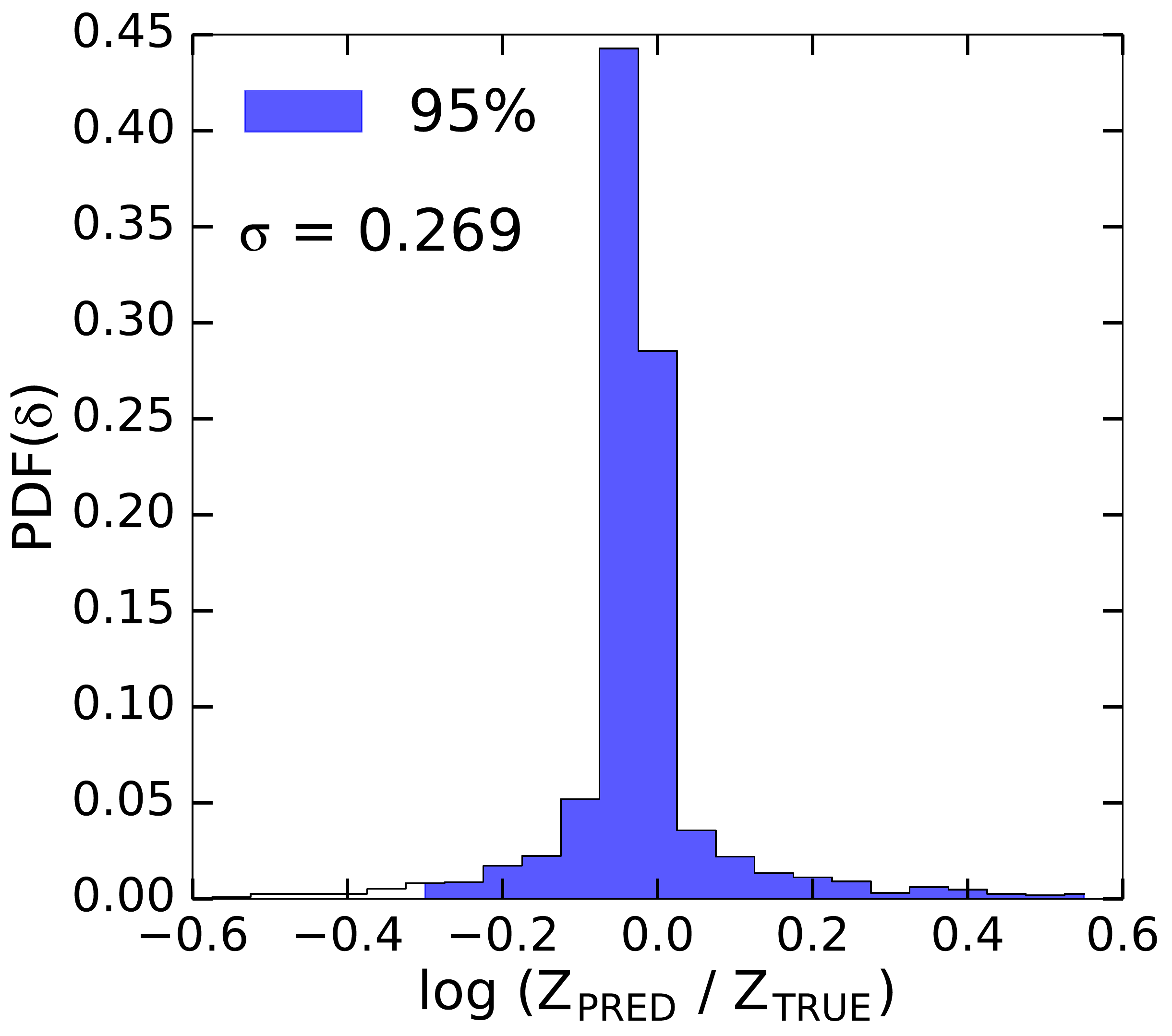}
	\includegraphics[width=0.45\textwidth]{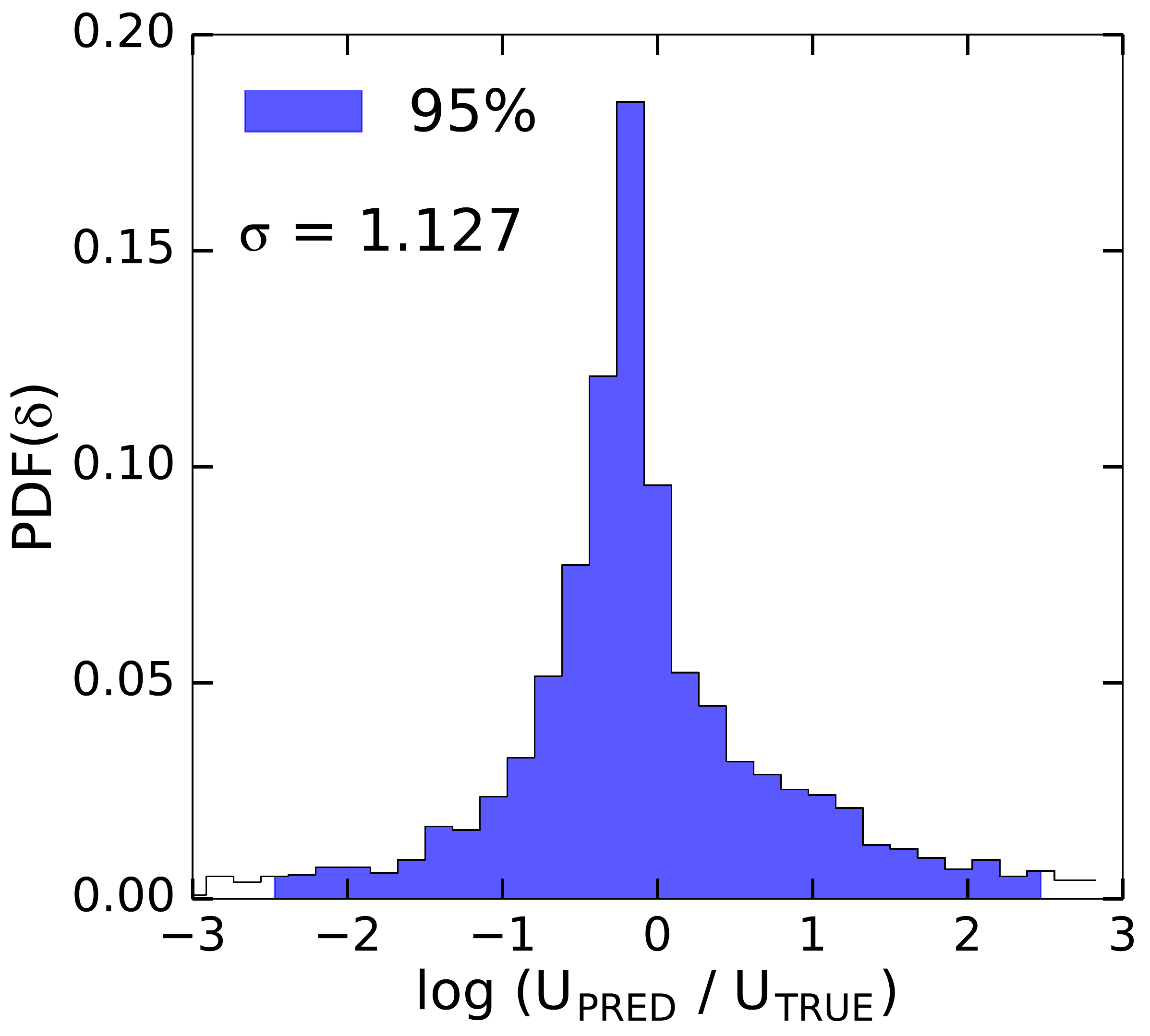}
	\caption{As in Fig. \ref{fig:predicted_vs_true_hist_sigma} but considering a line intensity threshold $f({\rm H}\alpha) = 1/50$ (see text for the details).}
	\label{fig:predicted_vs_true_hist_sigma}
\end{figure*}

The dataset used to train \textlcsc{GAME} consists in a library of $3\times 10^4$ models chosen by randomly selecting the values of the four physical parameters in the ranges reported in Table \ref{table:grid}. The dataset used for the test (i.e. to predict the labels) consists instead of a [test] sample of $3\times 10^3$ models, also randomly constructed. Thus, although the way of constructing this testing sample is the same used to construct the training dataset, the AdaBoost algorithm had never seen these objects before. 

The results of the predictive \textlcsc{GAME} tests performed by using the AdaBoost with Decision Tree as base learner are shown in Fig. \ref{fig:predicted_vs_true}. The fraction of models for which the actual (i.e. the known values used to generate the testing dataset) and predicted values deviate by a factor $>2$ are 14.8\% ($n$), 1.2\% ($Z$), 1.7\% ($N_H$), and 23.2\% ($U$). The lower-quality predictive performances are somewhat expected. In fact, the determination of the ionization parameter is particularly challenging: it involves the reconstruction from the emerging filtered spectrum of the original $U$ value at the source, that is highly degenerate with the $N_H$. \textlcsc{GAME} delivers top-quality predictions for $Z$ and $N_{H}$, which are almost perfectly recovered by the algorithm. 

A different way to appreciate \textlcsc{GAME} predictive performances is to look at the probability distribution function (PDF, Fig. \ref{fig:predicted_vs_true_hist}) of the fractional variation between predicted and true physical properties, defined as $\delta_X = \log (X_{\it PRED}/X_{\it TRUE})$. In each plot of Fig. \ref{fig:predicted_vs_true_hist}, the blue shaded area contains 95\% of the models. As previously mentioned, the best predictive performances are achieved for $N_H$ and $Z$. For example, for the column density, 95\% of the predicted values differ by $< 60$\% from the actual ones.

\subsection{Weak Lines}\label{sec:weak}
Up to now we have considered idealized synthetic spectra. They are idealized because of their \quotes{infinite SNR ratio}: they exhibit in fact weak emission lines that would be extremely hard to detect experimentally (e.g. up to $10^{5}$ times fainter than the H$\alpha$ or [OIII] lines). Hence, in our library of synthetic spectra on average 500 emission lines are available (i.e. with intensities different from zero). Although \textlcsc{GAME} does not use all these lines because some of them are unimportant for the construction of the decision trees, this idealized situation is unlikely to be reached in a typical observational setup.

We therefore investigated how \textlcsc{GAME} behaves when a \quotes{detection threshold} is added. We consider null the emission from lines whose intensity is less than a certain fraction f(H$\alpha$) of the H$\alpha$, therefore excluding them from the model. Formally, this approach is equivalent to consider a spectrum whose Signal to Noise Ratio (SNR) is SNR $= \sigma/f({\rm H}\alpha)$, where $\sigma$ is the r.m.s. noise. 

\begin{figure}
	\centering
	\includegraphics[width=0.9\linewidth]{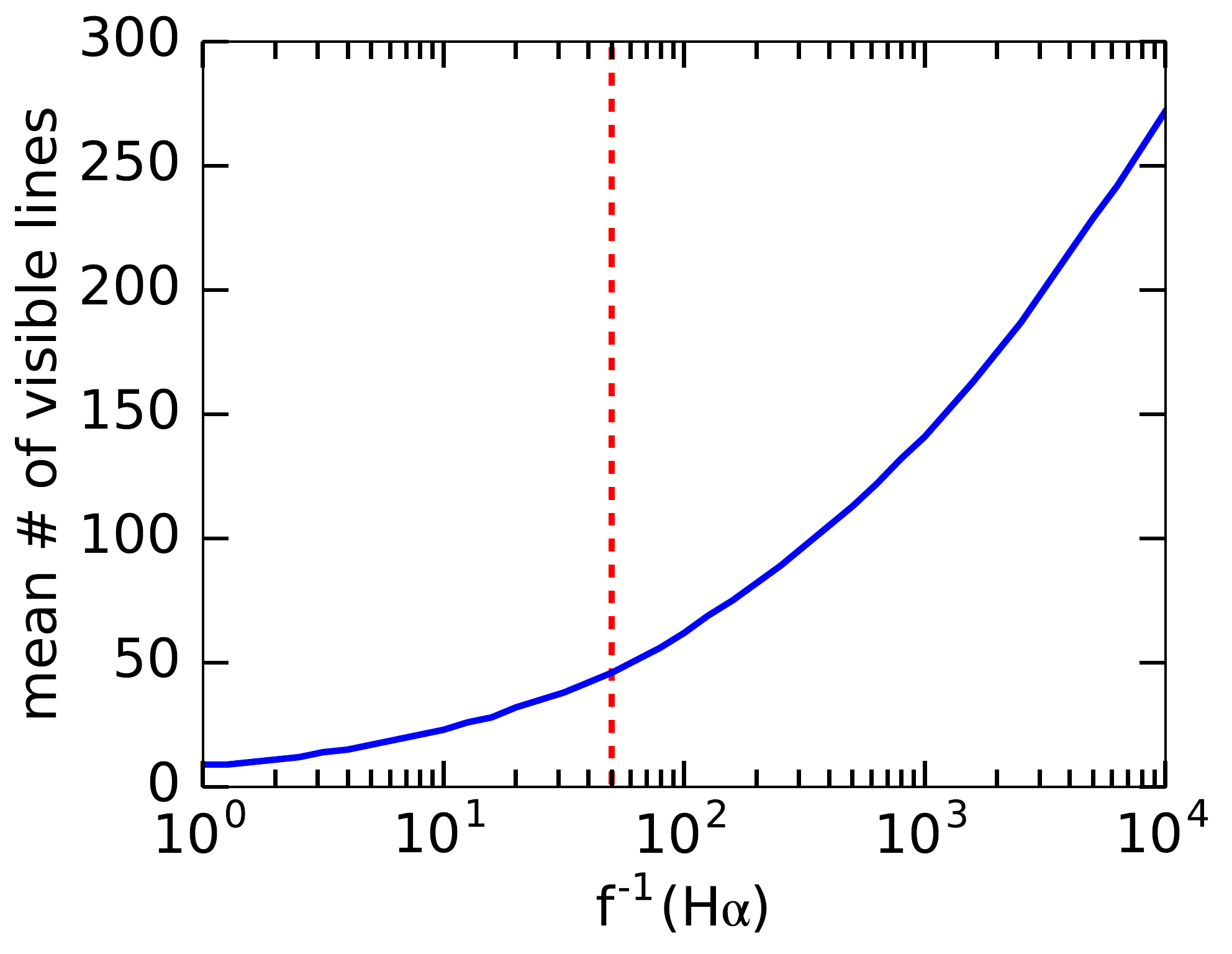}
	\caption{Mean number of available lines in our library of synthetic spectra as a function of $f^{-1}({\rm H}\alpha)$, the fraction of the H$\alpha$ line intensity used as a threshold. For $f({\rm H}\alpha) = 1/50$ (red dashed line) the mean number of available lines is 50.}
	\label{fig:soglia_linee}
\end{figure}

In Fig. \ref{fig:soglia_linee} we report the mean number of available lines in our grid of synthetic spectra as a function of $f^{-1}({\rm H}\alpha)$. The red dashed line in the figure is the threshold $f({\rm H}\alpha) = 50$ that we have adopted as a reference in this work. This value is easily reached with state of the art instruments, and it is typical of several spectra obtained in observations \citep{Cresci2015}. 

We construct two new datasets with spectra that contains only emission lines with an intensity higher than 1/50 of the H$\alpha$ line. One set is used for the training and the other for the testing phase.

Results of this approach are shown as scatter plots in Fig. \ref{fig:predicted_vs_true_sigma} and the resulting PDFs are shown in Fig. \ref{fig:predicted_vs_true_hist_sigma}. The fraction of models for which the \quotes{real} (i.e. the known values used to generate the testing dataset) and predicted values deviate by a factor $>2$ are are 21.9\% (density), 2.6\% (metallicity), 2.5\% (column density), and 27.3\% (ionization parameter). For $f({\rm H}\alpha) > 100$, results become similar to the ideal case.

We emphasize that \textlcsc{GAME} is easy to implement as well as extremely fast even on a laptop computer. Typical run times to train models using our library of $3\times10^4$ spectra is about 10 minutes for a single processor run. Therefore, accurately training \textlcsc{GAME} based on the SNR of the observed input spectra presents no difficulties.

\subsection{Sum of Different Phases}\label{sec:phases}
A line of sight (los) passing through a galaxy can cross different ISM phases, e.g. cold neutral medium (CNM), warm neutral medium (WNM), warm ionized medium (WIM) or a dense giant molecular cloud (GMC). The resulting spectrum is then the sum of the spectra from individual phases. This is the common case for observations, hence it is important to test \textlcsc{GAME} in these conditions. 

\begin{figure*}
	\centering
	\includegraphics[width=1.0\linewidth]{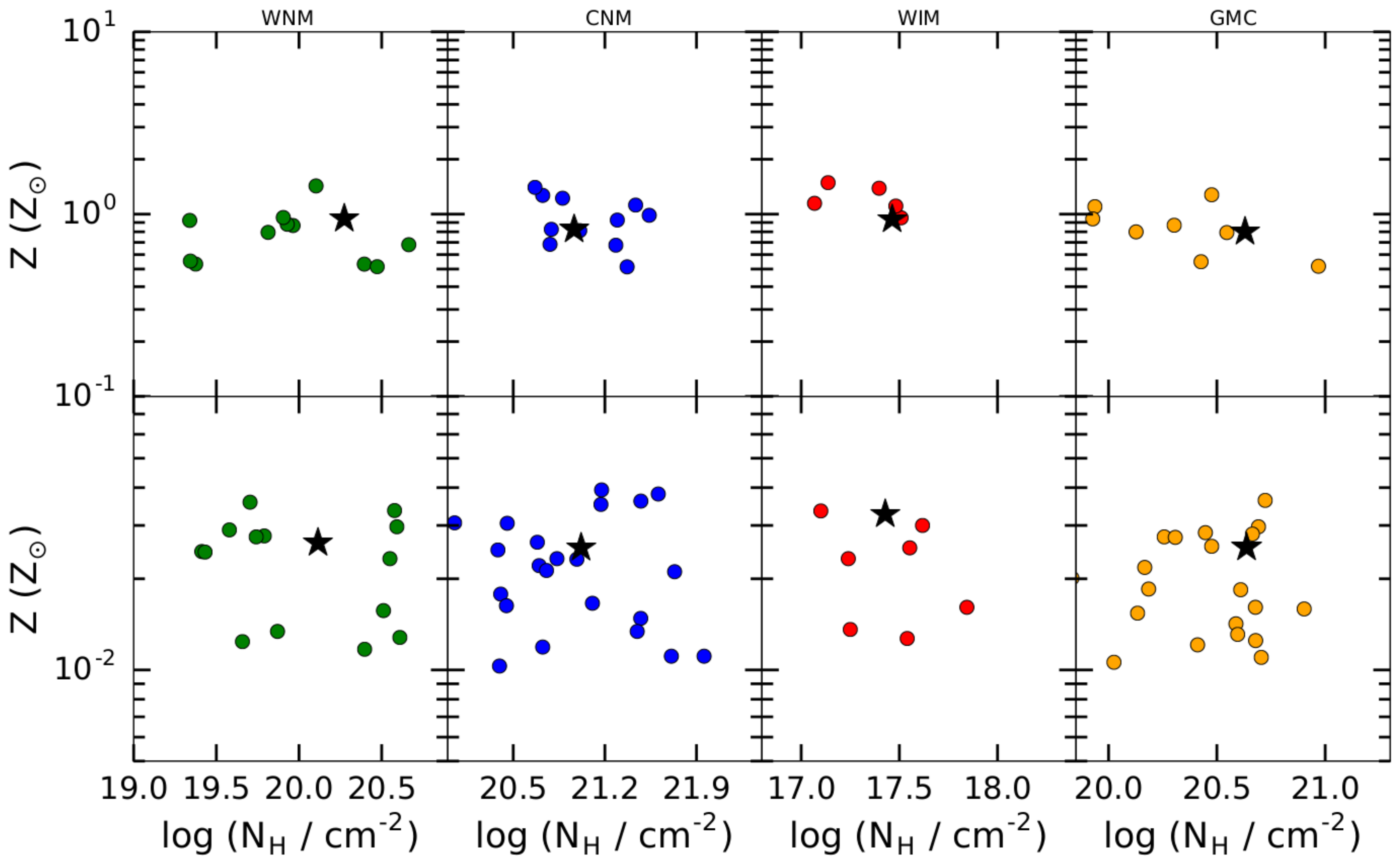}
	\caption{Metallicity vs column density for different models in our library representing different phases of the ISM: WNM (green), CNM (blue), WIM (red) and GMC (orange) computed for values of the metallicity $Z\simeq\zsun$ (upper panel) and $Z\simeq 0.02 \zsun$ (lower panel). The black stars represent the values inferred by \textlcsc{GAME} for the metallicity and column density using the sum of different spectra as the input.}
	\label{fig:somma_spettri}
\end{figure*}

To this aim, we first select spectra from the library labeled with parameters close to the typical values for ISM phases: WNM ($n \sim 1 \,\cc$,  $U \sim 10^{-4}$), CNM ($n\sim 10^{2}\cc$, $U \sim 10^{-4}$), WIM ($n\sim 10^{-2} \cc$,  $U\sim 10^{2}$) and GMC ($n\sim10^{3}\cc$, $U\sim 10^{-4}$). The size of the CNM, WNM and WIM phases is $l \sim 20$ pc, while for GMC we assumed a size of $l \sim 2$ pc \citep{Larson1981,Falgarone1992,Ossenkopf2002,Heyer2009}. Then we sum a variable number of spectra into the final one:

\begin{equation}
S^j(\lambda, Z, N_H) =  \sum_i S_i^j(\lambda, n^j, Z^j, N_H^j, U^j)\,,
\end{equation}
where $j$ labels the phase ($j=$ CNM, WNM, WIM, GMC) and $i=1,..., N$ is the $i$-th component along the line of sight. We have run cases with $Z\simeq 1 Z_\odot$ and $Z\simeq 0.02 Z_\odot$.

The comparison between the $Z,\, N_H$ values inferred by \textlcsc{GAME} for the final spectrum and the ones of the individual phases is shown in Fig. \ref{fig:somma_spettri}. \textlcsc{GAME} performs quite well recovering the emission-weighted values that are intermediate between the inputs of the individual components. This has been verified for all four phases and independently of the assigned mean $Z$.

\begin{figure}
	\centering
	\includegraphics[width=0.9\linewidth]{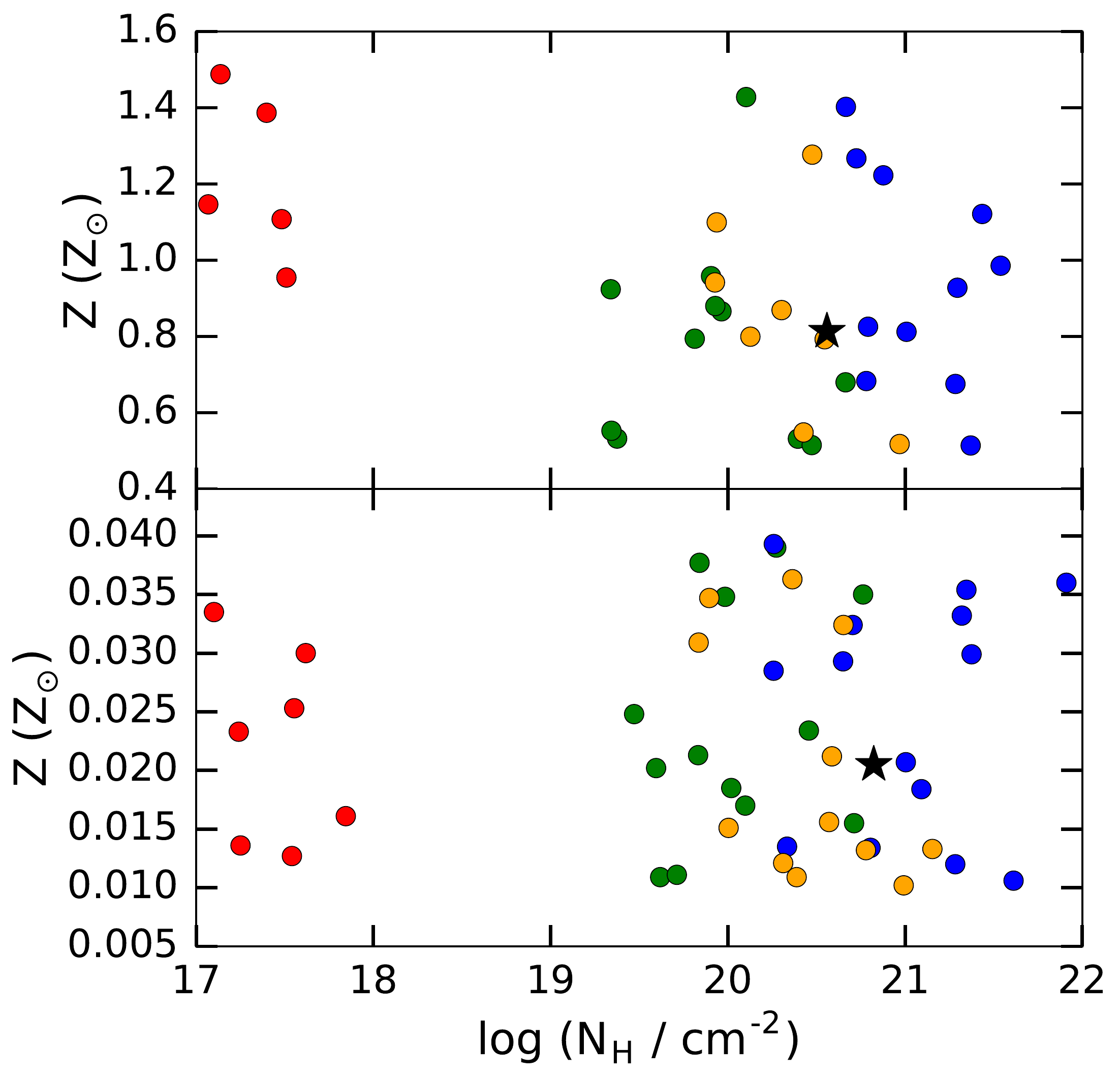}
	\caption{The inferred metallicity and column density (black stars) returned by \textlcsc{GAME} for a composite line of sight (Fig. \ref{fig:somma_spettri}): WNM (green), CNM (blue), WIM (red) and GMC (orange). The case for $Z\simeq \zsun$ and $Z\simeq 0.02\zsun$ are shown in the upper and lower panels, respectively.}
	\label{fig:somma_spettri_total}
\end{figure}

As a next step in complexity, we tested a spectrum that is a random combination of different phases along a los (Fig. \ref{fig:somma_spettri_total}). As \textlcsc{GAME} returns a single quadruple of ($n,\, Z,\, N_H,\, U$) values, the outcome is biased towards the phases characterized by a larger column density. In other words, \textlcsc{GAME} is most sensitive to the emitting phase with the largest gas mass.

Although this limitation, the outcome is very satisfactory as the inferred ($Z, N_H$) values are well within the range of the individual phases.

\subsection{Comparison with Calibrated Diagnostics}
We compare the calibration of two popular metallicity diagnostics, R$_{23}$ (see eqs. \ref{eq:r_23}) and [NII]$\lambda$6584 / H$\alpha$, with \textlcsc{GAME}, in order to evaluate the relative performances. As already mentioned, these two diagnostics are based on nebular emission lines and they have also been calibrated on HII regions. For the comparison to be meaningful, we have chosen in our library spectra with physical properties describing standard ionized nebulae: $T\sim 10^{4}$ K, ionized hydrogen fraction  $>90$\%, $N_{H}\lesssim 10^{20}\ccol$.

\begin{figure*}
	\centering
	\includegraphics[width=1.0\linewidth]{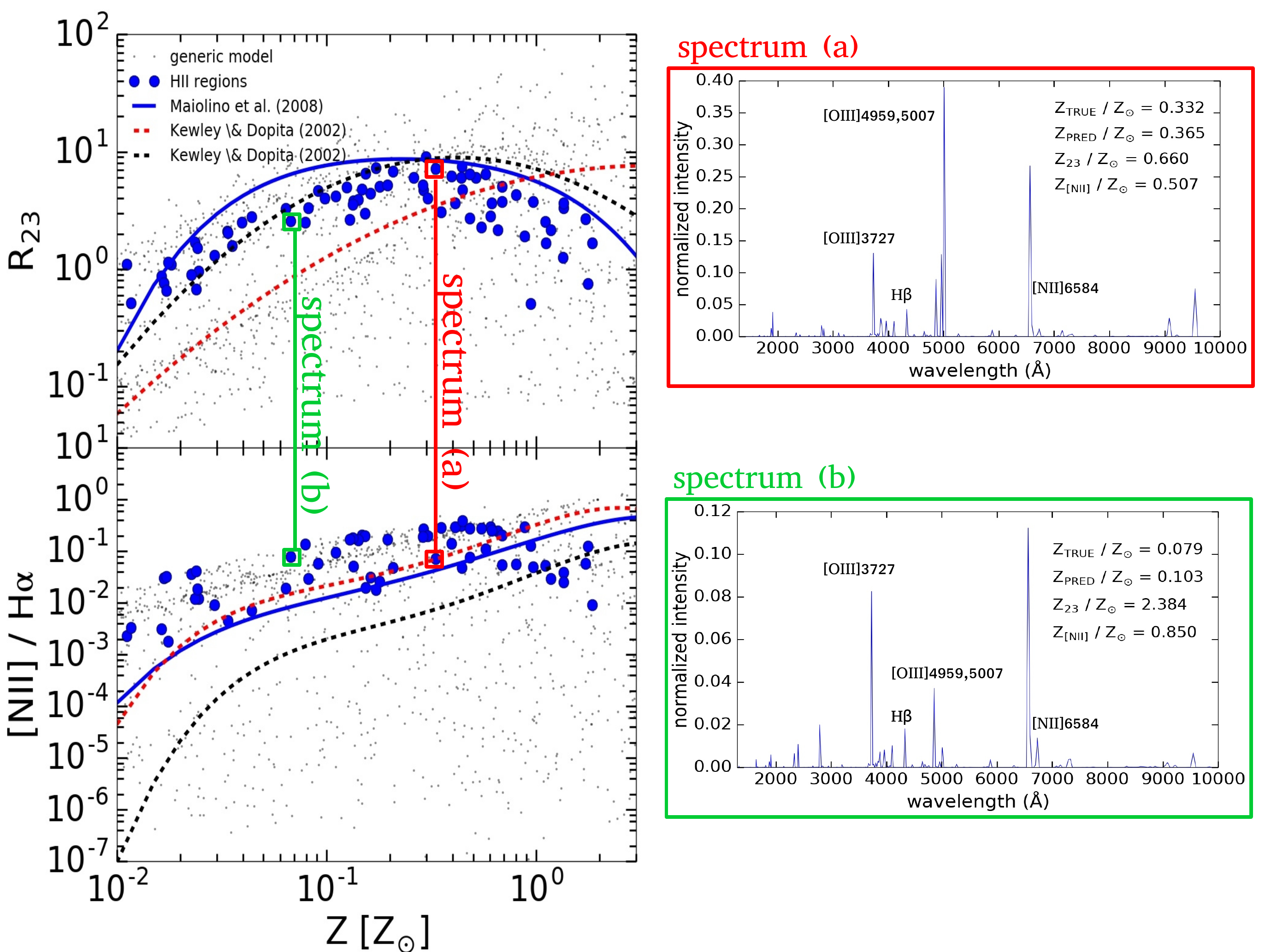}
	\caption{Relations between emission line ratios and gas metallicity. Green dots are generic models in our library, and blue circles are models representing HII regions. The blue solid line shows the empirical calibrations from \citet{Maiolino2008}. The red and black dashed lines are the theoretical calibrations reported in \citet{Kewley2002} respectively for a $U = 1.6 \cdot 10^{-4}$ and $U = 10^{-2}$, respectively. The red panel (a) shows a spectrum for which the calibrations give a value for the metallicity near to the true value used to generate the model. The green panel (b) shows instead a spectrum for which the calibrations are not good indicators for the metallicity.}
	\label{fig:confronto_diagnostici_total}
\end{figure*}

In Fig. \ref{fig:confronto_diagnostici_total} we show the values of R$_{23}$ and [NII]$\lambda$6584 / H$\alpha$ for generic models (gray dots), and those representing an HII region (blue circles) for $10^{-2} < Z / \zsun < 3$. The blue solid lines are the empirical calibrations given by \citet{Maiolino2008} for R$_{23}$: 

\begin{equation}
\begin{split}
\text{log}(R_{23}) = 0.7462 - 0.7149 x - 0.9401 x^2 \\
- 0.6154 x^3 - 0.2524 x^4
\end{split}
\end{equation}

and for [NII]$\lambda$6584 / H$\alpha$:

\begin{equation}
\begin{split}
\text{log}\left[ \frac{\text{F([NII] $\lambda$6584)}}{\text{F(H$\alpha$)}}\right] = -0.7732 + 1.2357 x\\
- 0.2811 x^2 - 0.7201 x^3 - 0.3330 x^4\,,
\end{split}
\end{equation}
where $x = \log(Z/\zsun) = 12+\log({\rm O/H}) - 8.69$ and $F$ are the reddening-corrected fluxes. As pointed by \citet{Maiolino2008}, the previous relations are strictly valid only in the range 7.0 $\lesssim 12+\log({\rm O/H}) \lesssim$ 9.2. Outside this metallicity range, the use of these relations relies on extrapolation. 

The dashed lines represent theoretical calibrations based on the grid of photoionization models provided by \citet{Kewley2002} for an ionization parameter $U = 1.6 \times 10^{-4}$ (red dashed lines in Fig. \ref{fig:confronto_diagnostici_total}):

\begin{equation}
\text{log}(R_{23}) = -27.0004 + 6.0391 y - 0.327006 y^2\,,
\end{equation}

\begin{equation}
\begin{split}
\text{log}\left[ \frac{\text{F([NII] $\lambda$6584)}}{\text{F(H$\alpha$)}}\right] = -2700.08 + 1335.14 y\\
- 247.533 y^2 + 20.3663 y^3 - 0.62692 y^4;
\end{split}
\end{equation}
and $U = 10^{-2}$ (black dashed lines in Fig. \ref{fig:confronto_diagnostici_total}):

\begin{equation}
\text{log}(R_{23}) = -45.6075 + 11.2074 y - 0.674460 y^2\,,
\end{equation}

\begin{equation}
\begin{split}
\text{log}\left[ \frac{\text{F([NII] $\lambda$6584)}}{\text{F(H$\alpha$)}}\right] = -3100.57 + 1501.77 y\\
- 272.883 * y^2 + 22.0132 y^3 - 0.6646 y^4\,,
\end{split}
\end{equation}
where $y = 12+\log({\rm O/H})$.

Some differences are present between our HII region models (blue circles) and the considered theoretical and empirical calibrations (solid and dotted lines). These are due to the different photoionization code used, the assumption of spherical vs plane-parallel geometry, isochoric vs isobaric assumptions for the gas. However, they are relatively minor.

We compute in this case $Z$ inferred from two different kinds of HII region models: (a) spectra for which the R$_{23}$ value is very close to the empirical calibration (blue solid line in Fig. \ref{fig:confronto_diagnostici_total}, taken as reference) and (b) spectra whose R$_{23}$ strongly differs from the calibration.

For case (a) (red panel of Fig. \ref{fig:confronto_diagnostici_total}) the \quotes{true} value (the one used to generate the spectrum) and the \textlcsc{GAME}-predicted one are $Z_{TRUE} = 0.332\, \zsun$ and $Z_{PRED} = 0.365 \zsun$; the difference is $< 10$\%. For the same case though, the metallicity inferred using calibrations based on the R$_{23}$ and [NII]$\lambda$6584 / H$\alpha$ diagnostic are instead $Z_{23} = 0.660 \zsun$, and $Z_{\rm [NII]} = 0.507 \zsun$, respectively. We see that the performance is much worse that the one delivered by \textlcsc{GAME}. In fact, the empirical methods predict values in excess by almost a factor of 2 with respect to the actual value.

For case (b) (green panel of Fig. \ref{fig:confronto_diagnostici_total}) the true and the predicted value are respectively $Z_{TRUE} = 0.0792 \zsun$ and $Z_{PRED} = 0.1027 \zsun$, i.e. still in good agreement. However, using the calibrations we get instead $Z_{23} = 2.384 \zsun$ and $Z_{\rm [NII]} = 0.850 \zsun$, i.e. they both largely overestimate $Z$ by more than one order of magnitude.

We conclude that \textlcsc{GAME} appears to be able to extract physical conditions of the gas in a much more precise and reliable way with respect to standard indicators.

Finally, we stress that for the spectra not arising from HII regions (green dots in Fig. \ref{fig:confronto_diagnostici_total}) the previous calibrators cannot be applied. This happens since it does not longer exist a correlation between the diagnostic value and $Z$. A SML approach, like the one implemented by \textlcsc{GAME}, is however capable to \textit{extract information from all detected lines}. This allows a range of applications that is not only restricted to HII regions, but can extend to a wide variety of ISM phases.

\subsection{Connecting theory and observations with GAME}
The most natural use of \textlcsc{GAME} is to extract the physical properties of galaxies from spectroscopic emission line observations of galaxies. The full power of the code is manifest when it is used in combination with spatially resolved spectroscopy, like the one obtained from Integral Field Units. From such data cubes it will be possible to readily and robustly obtain physical properties (density, metallicity, ionization parameter) that represent key information to understand galaxy evolution. As new instruments, like MUSE, and telescopes (JWST, ALMA, TMT, E-ELT) will be able to obtain for the first time this type of data also on high redshift galaxies, a more comprehensive approach to interpret the spectra, as the one presented here, is mandatory to completely exploit their power.   

\textlcsc{GAME} can be also applied to synthetic maps constructed from galaxy simulations. The aim of this procedure is twofold. On the one hand, it will make possible to perform controlled experiments in which the parameters derived using \textlcsc{GAME} on the synthetic spectra can be directly compared with the simulated ones from which the maps have been constructed. Such procedure allows to better understand uncertainties of the method and individuate possible critical points. Secondly, it gives the opportunity to blindly analyze real and simulated maps of a given object and quantify the level at which the various properties of a galaxy are reproduced by a numerical model.

Extensions of \textlcsc{GAME} to additional applications, as for example absorption line spectra, different spectral bands, inclusion of kinematic information and others can be easily envisaged. The SML algorithms on which \textlcsc{GAME} is based are flexible enough that they can be easily modified to applications suiting a wide range of data types and quality.

\section{Summary and discussion}
We have introduced a code called \textlcsc{GAME} (GAlaxy Machine learning for Emission lines) to infer key galaxy ISM properties as density ($n$), metallicity ($Z$), column density ($N_{H}$) and ionization parameter ($U$). \textlcsc{GAME} is based on the Supervised Machine Learning (SML) algorithm AdaBoost with Decision Trees as base learner.

Using a large library of synthetic spectra ($3\times 10^{4}$ models for the training phase + $3\times 10^3$ for the testing phase) we have shown that \textlcsc{GAME} delivers excellent predictive performances, especially for the estimates of $Z$ and $N_{H}$. A comparison with the most widely used emission line diagnostics showed that \textlcsc{GAME} can overcome well known calibration difficulties.

\textlcsc{GAME} retains optimal performances even when more than 80\% of the lines potentially present in the observed spectral range are discarded, e.g. because these might be too weak to be detected by a given observation. For example, predicted physical parameters, like $Z$, deviate by a factor $>2$ from the actual ones only in 2.5\% of the models. We also tested the ability of \textlcsc{GAME} to deal with complex, multi-phase lines of sight, obtaining very satisfactory answers (for details see Sec. \ref{sec:phases}).

We emphasize that \textlcsc{GAME} is easy to implement as well as extremely fast even on a laptop computer. Typical training times using our library of $3\times 10^{4}$ spectra are $\approx 10$ minutes and the time required to infer physical parameter values for a given input spectrum is only less than few seconds. Therefore, accurately training \textlcsc{GAME} to accommodate the specific SNR of the observed input spectra presents no difficulties. 

It is worthwhile to add some remarks concerning the comparison with currently used methods, based on emission line ratios. Different galaxy properties can result in almost equal line ratio for some of the lines. Thus, more line ratios are generally required to break this degeneracy.

The SML approach can overcome these difficulties because (a) it makes use of all the available information present in the spectrum simultaneously, meaning that it is not necessary to choose a priori a subset of lines to use; (b) the training phase is extremely fast, and the code can easily adapt to new conditions (e.g. a different SNR ratio of the spectrum, see Sec. \ref{sec:weak}). The most fundamental aspect is that without any calibration, once trained, the algorithm provides an estimate of the main physical properties with no degeneracy.

Furthermore, we can compare the SML method adopted in this work with the \quotesing{traditional} data fitting technique. For an excellent review on the two cited approaches (\quotesing{algorithmic} vs \quotesing{traditional}) we refer the interested reader to the work by \citet{Breiman2001}. There are advantages and disadvantages in both these techniques and one should consider the best suited for the science case under study. 

Both approaches are based on a strong underlying assumption: the model\footnote{For example, photoionization codes alternative to \textlcsc{CLOUDY} are \textlcsc{MAPPINGS} \citep{Sutherland1993,Allen2008} and \textlcsc{MOCASSIN} \citep{Ercolano2003,Ercolano2005}.}, the chosen physical properties range and the prescriptions used to generate the library do capture the essential physics governing the ISM. It must be stressed that, to get an accurate estimate of the ISM physical properties, one must explore the largest possible range of parameter values when producing the library (see Table \ref{table:grid}). This is not always easy and the resulting grid from all the possible combinations of these values can be very large. Although model fitting techniques or Bayesian approaches \citep{Blanc2015} are very powerful, they suffer from some limitations. The best way to constrain a particular model is in fact to use as many observational constraints as possible. For a Bayesian approach this can be a problem, because using hundreds of features at one time is extremely time consuming. Moreover, adapting a code to deal with an observational spectrum with different wavelength range or with a different SNR ratio can be computationally very expensive.

In this context, an important advantage of the SML method with respect to the Bayesian one is that its performance is not affected by the finite number of models within the library used during the training. In fact, the SML technique, as suggested by the name itself, is capable to "learn" and explore hidden patterns within the library parameter space. In other words, the SML method is not limited to "recover" parameter values included in the library, but it can also "predict" results that are not part of the original one.

Moreover, \textlcsc{GAME} can be effectively coupled to the analysis of Integral Field Unit (IFU) spectroscopic data, and synthetic data from numerical galaxy simulations. These applications will be demonstrated in a forthcoming study. 

We finally point out that, in addition to HII regions, \textlcsc{GAME} allows to infer the physical properties of Photo Dissociation Regions. These are now being resolved in nearby galaxies. With JWST and ALMA we will be able to obtain comparable results also for high redshift systems.

\section*{Acknowledgements}
Acknowledgments: We thank D. Cormier, L. Vallini and S. Viti for useful discussions, and K. Volk for help with \textlcsc{CLOUDY}. We thank also the referee G. Ferland for insigthful comments. This research was supported in part by the National Science Foundation under Grant No. NSF PHY11-25915.

\bibliographystyle{mnras}
\bibliography{bibliography}

\end{document}